\documentclass[nohyper]{JHEP3}

\usepackage{multirow}
\usepackage{fix-cm}
\usepackage{cite}
\usepackage{epsfig}
\usepackage{amsmath}
\usepackage{graphicx}
\usepackage{dcolumn}
\usepackage{bm}

\def\iscol#1#2{{\buildrel#1\parallel#2\over\longrightarrow}}
\def\beq{\begin{equation}}
\def\eeq{\end{equation}}
\def\beqa{\begin{eqnarray}}
\def\eeqa{\end{eqnarray}}

\def \as {\relax\ifmmode\alpha_s\else{$\alpha_s${ }}\fi}
\def\e{\epsilon}
\def\NeqFour{{\cal N}=4}
\def\dip{{\rm dip}}
\def\kin{{\rm kin}}

\newcommand{\lsim}{
\mathrel{\hbox{\rlap{\hbox{\lower4pt\hbox{$\sim$}}}\hbox{$<$}}}}
\newcommand{\gsim}{
\mathrel{\hbox{\rlap{\hbox{\lower4pt\hbox{$\sim$}}}\hbox{$>$}}}}
\newif\ifdraft
\drafttrue
\newif\ifpreprint
\preprinttrue

\def\sect#1{Sec.~{\ref{#1}}}
\def\fig#1{fig.~{\ref{#1}}}

\def\eqn#1{eq.~(\ref{#1})}
\def\Eqn#1{Equation~(\ref{#1})}
\def\eqns#1#2{eqs.~(\ref{#1}) and~(\ref{#2})}

\def\lr{\leftrightarrow}

\newcommand{\JHEP}[1]{JHEP {\bf #1}}

\psfull
\preprint{SLAC--PUB--13816\\
Edinburgh 08/2009\\
DFTT 69/2009}

\title{On soft singularities at three loops and beyond}

\author{Lance J. Dixon\\
SLAC National Accelerator Laboratory, Stanford University,\\
Stanford, CA 94309, USA}
\author{Einan Gardi \\
School of Physics, The University of Edinburgh\\
Edinburgh EH9 3JZ, Scotland, UK}
\author{Lorenzo Magnea \\
Dipartimento di Fisica Teorica, Universit{\`a} di Torino, and\\
INFN, Sezione di Torino, Via P. Giuria 1, I-10125 Torino, Italy}

\abstract{
We report on further progress in understanding soft singularities 
of massless gauge theory scattering amplitudes.  Recently, a 
set of equations was derived based on Sudakov factorization,
constraining the soft anomalous dimension matrix of multi-leg
scattering amplitudes to any loop order, and relating it to the cusp
anomalous dimension. The minimal solution to these equations was 
shown to be a sum over color dipoles.  Here we explore potential
contributions to the soft anomalous dimension that go beyond the
sum-over-dipoles formula.  Such contributions are constrained by
factorization and invariance under rescaling of parton momenta
to be functions of conformally invariant cross ratios.  Therefore,
they must correlate the color and kinematic degrees of freedom of
at least four hard partons, corresponding to gluon webs that connect 
four eikonal lines, which first appear at three loops.  We analyze 
potential contributions, combining all available constraints, including 
Bose symmetry, the expected degree of transcendentality, and the 
singularity structure in the limit where two hard partons become collinear.
We find that if the kinematic dependence is solely through products of 
logarithms of cross ratios, then at three loops there is a unique function 
that is consistent with all available constraints. If polylogarithms are 
allowed to appear as well, then at least two additional structures are 
consistent with the available constraints.  
}

\begin{document}

\section{Introduction}

Understanding the structure of gauge theory scattering amplitudes is
important from both the fundamental field-theoretic perspective, and
the pragmatic one of collider phenomenology. Infrared singularities, in 
particular, open a window into the all-order structure of perturbation 
theory and the relation between the weak and strong coupling limits;
at the same time, they provide the key to resummation of large logarithms 
in a variety of phenomenological applications.

The study of infrared singularities in QCD amplitudes, which has a
three-decade-long history~\cite{%
Mueller:1979ih,Collins:1980ih,Sen:1981sd,Polyakov:1980ca,%
Dotsenko:1979wb,Brandt:1981kf,StermanWeb,%
Brandt:1982gz,Sen:1982bt,Gatheral:1983cz,Bassetto:1984ik,%
Frenkel:1984pz,Collins:1981uk,Korchemsky:1985xj,Sterman:1986aj,%
Ivanov:1985np,Korchemsky:1987wg,Korchemsky:1988hd,%
Korchemsky:1988si,Collins:1989gx,%
Sterman:1995fz,Catani:1996jh,Contopanagos:1996nh,%
KOS,KOSjet,BSZ,Laenen:2004pm,Dokshitzer:2005ig,%
Magnea:2008ga,Korchemskaya:1994qp,Botts:1989kf,Kidonakis:1997gm,%
Magnea:1990zb,Magnea:2000ss,Sterman:2002qn,Catani:1998bh,%
Aybat:2006wq,Aybat:2006mz,Dixon:2008gr,Eynck:2003fn},
recently received a major
boost~\cite{Becher:2009cu,Gardi:2009qi,Becher:2009qa}. 
The factorization properties of soft and collinear modes, also referred 
to as Sudakov factorization, were combined with the symmetry of 
soft-gluon interactions under rescaling of hard parton momenta, and 
were shown to constrain the structure of singularities of any massless
gauge theory amplitude, to any loop order, and for a general number
of colors $N_c$.
A remarkably simple structure emerges as the simplest solution to 
these constraints.  All non-collinear soft singularities are generated
by an anomalous dimension matrix in color
space~\cite{Brandt:1982gz,Sen:1982bt,Korchemsky:1985xj,Ivanov:1985np,%
Botts:1989kf,KOS,KOSjet,BSZ,Dokshitzer:2005ig}.  In the simplest solution, 
this matrix takes the form of a sum over color dipoles, corresponding to 
pairwise interactions among hard partons.  This interaction is governed by 
a single function of the strong coupling, the cusp anomalous dimension, 
$\gamma_K(\alpha_s)$. The simplicity of this result is remarkable, 
especially given the complexity of multi-leg amplitude computations 
beyond tree level. The color dipole structure of soft singularities appears 
naturally at the one-loop order~\cite{KOS,KOSjet,Catani:1996jh,Catani:1998bh},
where the interaction is genuinely of the form of a single gluon exchange 
between any two hard partons. The validity of this structure at two loops
was not obvious {\it a priori}; it was discovered through the explicit 
computation of the anomalous dimension
matrix~\cite{Aybat:2006wq,Aybat:2006mz}.
 
This remarkable simplicity is peculiar to the case of massless gauge
theories: recent work~\cite{Kidonakis:2009ev,Mitov:2009sv,Becher:2009kw,%
Beneke:2009rj,Czakon:2009zw,Ferroglia:2009ep,Ferroglia:2009ii,Kidonakis:2009zc}
has shown that the two-loop matrix, when at least two colored legs are
massive, is not proportional to the one-loop matrix, except in particular
kinematic regions. In general, in the massive case, there are new
contributions that correlate the color and momentum degrees of freedom of
at least three partons, starting at two loops.  These contributions vanish
as ${\cal O}(m^4/s^2)$ in the small mass
limit~\cite{Ferroglia:2009ep,Ferroglia:2009ii}.

Given that all existing massless results are consistent with the
sum-over-dipoles formula, it is tempting to conjecture that it gives the full
answer~\cite{Bern:2008pv,Becher:2009cu,Gardi:2009qi,Becher:2009qa,LaThuile}. 
As emphasized in Ref.~\cite{Gardi:2009qi}, however, constraints based on 
Sudakov factorization and momentum rescaling
alone are not sufficient to determine uniquely the 
form of the soft anomalous dimension. A logical possibility exists that 
further contributions will show up at the multi-loop level, which 
directly correlate the kinematic and color degrees of freedom of more 
than two hard partons. It is very interesting to establish whether these 
corrections exist, and, if they do not, to gain a complete understanding 
of the underlying reason. Beyond the significance of the soft singularities
themselves, a complete understanding of their structure may shed light
on the structure of the finite parts of scattering amplitudes.

Ref.~\cite{Gardi:2009qi} showed that precisely two classes of
contributions may appear as corrections to the sum-over-dipoles
formula. The first class stems from the fact that the sum-over-dipoles
formula provides a solution to the factorization-based constraints only
if the cusp anomalous dimension, $\gamma_K^{(i)} (\alpha_s)$, 
associated with a hard parton in representation $i$ of the gauge group,
obeys $\gamma_K^{(i)}(\alpha_s) = C_i \, \widehat{\gamma}_K 
(\alpha_s)$, where $\widehat{\gamma}_K$ is universal and $C_i$ is the
quadratic Casimir of representation $i$, $C_i = C_A$ or $C_F$ for gluons 
and quarks, respectively. This property is referred to as `Casimir scaling' 
henceforth. Casimir scaling holds through three
loops~\cite{Moch:2004pa,Kotikov:2004er}; an interesting open question 
is whether it holds at four loops and beyond~\cite{AldayMaldacena}.
At four loops, the quartic Casimir first appears in the color factors
of diagrams for the cusp anomalous dimension. 
(In QCD, with gluons in the adjoint representation $A$, fermions in the
fundamental representation $F$, and a Wilson line in representation $R$,
the relevant quartic Casimirs are $d_A^{abcd}d_R^{abcd}$
and $d_F^{abcd}d_R^{abcd}$, where $d_X^{abcd}$ are totally symmetric
tensors in the adjoint indices $a,b,c,d$.)
However, Ref.~\cite{Becher:2009qa} provided some
arguments, based on factorization and collinear limits of 
multi-leg amplitudes, suggesting that Casimir scaling might actually hold at
four loops. In the strong coupling limit, it is known 
to break down for $\NeqFour$ super-Yang-Mills theory in the large-$N_c$ 
limit~\cite{Armoni}, at least when $\gamma_K$ is computed
for Wilson lines in a special class of representations of the gauge group.

The second class of corrections, the one on which we focus here, can
occur even if the cusp anomalous dimension obeys Casimir scaling.
In this case, the sum-over-dipoles formula solves a set of inhomogeneous linear
differential equations, which follow from the constraints of Sudakov
factorization and momentum rescalings.  However, we can contemplate 
adding solutions to the homogeneous differential equations,
which are provided by arbitrary functions of conformally (and rescaling) 
invariant cross ratios built from the momenta of four hard
partons~\cite{Gardi:2009qi}.  Thus any additional terms must correlate
directly the momenta, and colors, of four legs.
Due to the non-Abelian exponentiation
theorem~\cite{StermanWeb,Gatheral:1983cz,Frenkel:1984pz} such 
contributions must originate in webs that connect four hard partons, 
which first appear at three loops.  From this perspective then, the absence 
of new correlations at two loops~\cite{Aybat:2006wq,Aybat:2006mz}, 
or in three-loop diagrams involving matter fields~\cite{Dixon:2009gx},
is not surprising, and it does not provide substantial new evidence in favor
of the minimal, sum-over-dipoles solution. The first genuine test is
from the matter-independent terms at three loops.  At this order,
purely gluonic webs may connect four hard partons, possibly inducing
new types of soft singularities that correlate the color and
kinematic variables of the four partons.

The most recent step in addressing this issue was taken in
Ref.~\cite{Becher:2009qa}, in which an additional strong constraint on the
singularity structure of the amplitude was established, based on the
properties of amplitudes as two partons become collinear.  Recall that the
primary object under consideration is the fixed-angle scattering
amplitude, in which
all ratios of kinematic invariants are taken to be of order 
unity. This fixed-angle limit is violated upon considering the special
kinematic situation where two of the hard partons become collinear.
An additional class of singularities, characterized by the vanishing invariant 
mass of the two partons, arises in this limit. The splitting amplitude is 
defined to capture this class of singularities.  It relates an $n$-parton
amplitude with two collinear partons to an $(n-1)$-parton amplitude, where 
one of the legs carries the total momentum and color charge of the 
two collinear partons.  The basic, universal property of the splitting
amplitude is that it depends only on the momentum and color degrees of
freedom of the collinear partons, and not on the rest of the process.

Splitting amplitudes have been explicitly computed, or extracted 
from known scattering amplitudes, at 
one~\cite{Bern:1994zx,Bern:1998sc,Bern:1999ry,Kosower:1999rx}
and two~\cite{Bern:2004cz,Badger:2004uk} loops.
A derivation of splitting-amplitude universality to all loop orders,
based on unitarity, has been given in the large-$N_c$
limit~\cite{Kosower:1999xi}.  The light-cone-gauge method for
computing two-loop splitting amplitudes~\cite{Bern:2004cz},
in which only the two collinear legs and one off-shell parton appear,
strongly suggests that the same all-orders universality extends
to arbitrary color configurations, not just planar ones.

Based on splitting-amplitude universality, Ref.~\cite{Becher:2009qa} 
established additional constraints on the singularity structure of amplitudes.
Using these constraints in conjunction with the Sudakov factorization
constraints discussed above, that paper excluded any possible
three-loop corrections depending linearly on logarithms of cross 
ratios. The final conclusion was, however, that more general functions
of conformal cross ratios that vanish in all collinear limits could not
be ruled out.

In the present paper we re-examine the structure of soft singularities at 
three loops. We put together all available constraints, starting with the 
Sudakov factorization constraints and Bose symmetry, and including 
the properties of the splitting amplitude and the expected degree of
transcendentality of the functions involved\footnote{Transcendentality 
here refers to the assignment of an additive integer $\tau$ for each type 
of factor in a given term arising in an amplitude or an anomalous dimension:
$\tau = 0$ for rational functions, $\tau = 1$ for factors of $\pi$ or single
logarithms, $\ln x$;  $\tau = n$ for factors of $\zeta(n)$, $\ln^n x$
or ${\rm Li}_n(x)$, {\it etc.}~\cite{Kotikov:2002ab}. We will provide more
examples in \sect{sec:maxtran}.}.
We make some plausible assumptions on the kinematic dependence,
and consider all possible 
products of logarithms, and eventually also polylogarithms. We find that 
potential contributions beyond the sum-over-dipoles formula are still possible
at three loops, but their functional form is severely constrained.  

The paper is organized as follows. We begin with three short 
sections in which we review the main relevant results of 
Refs.~\cite{Gardi:2009qi,Becher:2009qa}.  In \sect{sec:factorization} 
we briefly summarize the Sudakov factorization of the amplitude 
and the constraints imposed on the soft anomalous
dimension matrix by rescaling invariance of Wilson lines.
In \sect{sec:amplitude_ansatz} we present the sum-over-dipoles 
formula, the simplest possible solution to 
these constraints. In \sect{sec:SA} we review the splitting amplitude 
constraint.  The main part of our study is \sect{sec:corrections}, in 
which we put together all available constraints and analyze the possible 
color and kinematic structures that may appear beyond the 
sum-over-dipoles formula.  Most of the discussion is general, and 
applies to any loop order, but specific analysis is devoted to potential 
three-loop corrections. At the end of the section
we make a few comments concerning four-loop corrections. 
Our discussion throughout \sect{sec:corrections} focuses on amplitudes 
involving four colored partons, plus any number of color-singlet particles.
The generalization to the multi-parton case is presented in \sect{sec:n-leg}.  
Our conclusions are summarized in \sect{sec:conc}, while an
appendix discusses the special case of four-parton scattering at
three loops.


\section{Sudakov factorization and its consequences~\label{sec:factorization}}

We summarize here the infrared and collinear factorization properties 
of fixed-angle scattering amplitudes ${\cal M} \left(p_i/\mu, 
\alpha_s (\mu^2), \e \right)$ involving $n$ massless partons, 
plus any number of color-singlet particles, evaluated 
in dimensional regularization with $D = 4 - 2 \e$.  
We refer the reader to Ref.~\cite{Gardi:2009qi} for technical 
details and operator definitions of the various functions involved. 
Multi-parton fixed-angle amplitudes can be expressed in terms of their
color components ${\cal M}_L$ in a chosen basis in the vector
space of available color structures for the scattering process 
at hand. All infrared and collinear singularities of ${\cal M}_L$ can be
factorized~\cite{Sen:1982bt,Collins:1989gx,Sterman:1995fz,%
Sterman:2002qn,Aybat:2006mz,Dixon:2008gr,Gardi:2009qi} into jet 
functions $J_i$, one for each external leg $i$, multiplied by a 
(reduced) soft matrix $\overline{\cal S}_{LM}$,
\begin{eqnarray}
\label{facamp_bar}
  {\cal M}_{L} \left(p_i/\mu, \alpha_s (\mu^2),
  \epsilon \right) & = & 
  \overline{\cal S}_{L M} \left(\rho_{ij} , \alpha_s (\mu^2), \epsilon 
  \right) \,  H_{M} \left( \frac{2 p_i \cdot p_j}{\mu^2},
  \frac{(2 p_i \cdot n_i)^2}{n_i^2 \mu^2}, \alpha_s (\mu^2), 
  \epsilon \right)
  \nonumber \\ &&\hspace*{100pt} \times
  \prod_{i = 1}^n  
  J_i\left(  \frac{(2 p_i \cdot n_i)^2}{n_i^2 \mu^2},
  \alpha_s (\mu^2), \epsilon \right)
    \,\,,
\end{eqnarray}
leaving behind a vector of hard functions $H_M$, which are finite 
as $\e \to 0$. A sum over $M$ is implied. The hard
momenta\footnote{In our convention momentum conservation reads
$q+\sum_{i=1}^n p_i=0$, where $q$ is the recoil momentum carried
by colorless particles.} $p_i$ defining the amplitude ${\cal M}$ are 
assumed to be light-like, $p_i^2 = 0$, while the $n_i$ are 
auxiliary vectors used to define the jets in a gauge-invariant way, 
and they are not light-like, $n_i^2 \neq 0$.  
The reduced soft matrix $\overline{\cal S}_{LM}$ can be computed 
from the expectation value of a product of eikonal lines, 
or Wilson lines, oriented along the hard parton momenta,
dividing the result by $n$ eikonal jet functions 
${\cal J}_i$, which remove collinear divergences and leave only singularities 
from soft, wide-angle virtual gluons.  It is convenient to 
express the color structure of the soft matrix $\overline{\cal S}$ in a 
basis-independent way, in terms of operators ${\bf T}_i^a$, 
$a=1,2,\ldots,N_c^2 - 1$, representing the generators of ${\rm SU}(N_c)$ 
acting on the color of parton $i$ ($i=1,2,\ldots,n$)~\cite{Catani:1996jh}.
 
The partonic (quark or gluon) jet function solves two evolution
equations simultanously, one in the factorization scale $\mu$ and
another in the kinematic variable $(2p_i\cdot n_i)^2/n_i^2$
(see {\it e.g.} Ref.~\cite{Gardi:2009qi}). The latter equation
generalizes the evolution of the renormalization-group invariant
form factor~\cite{Magnea:1990zb}.  The resulting solution to
these equations can be written as~\cite{LaThuile}
\begin{eqnarray}
&& \hspace{-2mm}
J_i\left(\frac{(2p_i\cdot n_i)^2}{n_i^2}, \alpha_s(\mu^2), \epsilon 
\right)  = \,  H_{J_i} \left(1,
\alpha_s \left({\textstyle\frac{(2 p_i\cdot n_i)^2}{n_i^2}} \right),
\epsilon\right) \exp \Bigg\{ 
- \frac12 \int_{0}^{\mu^2} \frac{d\lambda^2}{\lambda^2} 
\gamma_{J_i} \left( \alpha_s(\lambda^2,\epsilon) \right) 
\nonumber \\
&&  \hspace{-2mm} + \, \frac{{\bf T}_i \cdot {\bf T}_i }{2} 
\int_0^{(2 p_i\cdot n_i)^2/n_i^2} \frac{d \lambda^2}{\lambda^2} 
\Bigg[ \frac14 \widehat{\gamma}_K 
\left( \alpha_s( \lambda^2, \epsilon) \right)
\ln \left( \frac{n_i^2 \lambda^2}{(2 p_i\cdot n_i)^2} \right)
+  \frac12 \widehat{\delta}_{{\overline{\cal S}}} \left(\alpha_s 
(\lambda^2, \epsilon) \right) \Bigg] \Bigg\} \, ,
\label{J_explicit}
\end{eqnarray}
where $H_{J_i}$ is a finite coefficient function, and all
singularities are generated by the exponent.
The solution depends on just three anomalous dimensions,
which are functions of the $D$-dimensional coupling alone:
$\gamma_{J_i}$ is the anomalous 
dimension of the quark or gluon field defining the jet
(corresponding to the quantity $\gamma^i$ defined
in Refs.~\cite{Becher:2009qa,Becher:2009cu}),
while
$\widehat{\gamma}_K=2\alpha_s/\pi+\cdots$ and
$\widehat{\delta}_{{\overline{\cal S}}}=\alpha_s/\pi+\cdots$ are,
respectively, the cusp anomalous dimension and an additional eikonal
anomalous dimension defined in Sec.~4.1 of Ref.~\cite{Gardi:2009qi}.
In eq.~(\ref{J_explicit}) we have already assumed that the latter
two quantities admit Casimir scaling, and we have factored out the
quadratic Casimir operator $C_i \equiv {\bf T}_i \cdot {\bf T}_i$.

Our main interest here is the reduced soft matrix
$\overline{{\cal S}}$, which takes
into account non-collinear soft radiation. It is defined entirely in
terms of vacuum correlators of operators composed of semi-infinite
Wilson lines (see {\it e.g.} Ref.~\cite{Gardi:2009qi}),
and depends on the kinematic variables
\begin{equation}
\rho_{ij} \, \equiv \, \frac{\left(-  \beta_i \cdot \beta_j \right)^2}
{\displaystyle \frac{2(\beta_i \cdot n_i)^2}{n_i^2}
\frac{2(\beta_j \cdot n_j)^2}{n_j^2}} \, = \,
\frac{  
\, \left| \beta_i \cdot \beta_j \right|^2 \, 
{\rm e}^{-2 {\rm i} \pi \lambda_{ij}} }
{\displaystyle \frac{2(\beta_i \cdot n_i)^2}{n_i^2}
\frac{2(\beta_j \cdot n_j)^2}{n_j^2}} \,  ,
\label{rhoij}
\end{equation}
which are invariant with respect to rescaling of all the Wilson line
velocities $\beta_i$.  The $\beta_i$ are related to the external momenta 
by $p_i^\mu=(Q/\sqrt{2})\beta_i^\mu$, where $Q$ is a hard scale whose
precise value will not be important here. The phases $\lambda_{ij}$ are
defined by $\beta_i \cdot \beta_j = - | \beta_i \cdot \beta_j |
{\rm e}^{-{\rm i} \pi \lambda_{ij}}$, where $\lambda_{ij} = 1$ if
$i$ and $j$ are both initial-state partons, or both final-state partons,
and  $\lambda_{ij}=0$ otherwise. Note that the sign of the phase in 
${\rm e}^{-{\rm i} \pi \lambda_{ij}}$ is determined by the
$+{\rm i}\varepsilon$ prescription for the Feynman propagator. 

The reduced soft matrix obeys the renormalization group equation
\begin{equation}
\mu  \frac{d}{d \mu} \overline{{\cal S}}_{L M}
\left( \rho_{i j}, \alpha_s, \epsilon \right) = - \,\sum_{N} 
\overline{{\cal S}}_{L N}\left( \rho_{i j}, \alpha_s, \epsilon \right) 
\,\,
\Gamma^{\overline{{\cal S}}}_{N M} \left( \rho_{i j}, 
\alpha_s \right) 
\,.
\label{rencalS} 
\end{equation}
The soft anomalous dimension matrix $\Gamma^{{\overline{\cal S}}}_{N M} \left( 
\rho_{i j}, \alpha_s \right)$, in turn, obeys the equation~\cite{Gardi:2009qi}
\begin{equation}
\label{constraints}
\sum_{j \neq i} \frac{\partial}{\partial \ln\rho_{ij}} \,
\Gamma^{{\overline{\cal S}}}_{N M} \left( 
\rho_{i j}, \alpha_s \right)  =  \frac{1}{4} \, \gamma_K^{(i)} 
\left( \alpha_s \right) \,\delta_{N M}\,, \qquad \qquad \forall\, i,\,N,M \,,
\end{equation}
found by considering a rescaling of the eikonal velocity $\beta_i$.
The simplest solution of this equation is the sum-over-dipoles
formula~\cite{Gardi:2009qi},
\begin{equation}
\Gamma^{\overline{\cal S}}_{\dip} \left( \rho_{i j}, \alpha_s \right) =
- \, \frac{1}{8} \, \widehat{\gamma}_K(\alpha_s)
\sum_{i = 1}^n \sum_{j\neq i} \, \ln \rho_{ij} \,
{\bf T}_i \cdot {\bf T}_j
+ \frac{1}{2} \, \widehat{\delta}_{{\overline{\cal S}}}(\alpha_s) \,
\sum_{i = 1}^n \, {\bf T}_i \cdot {\bf T}_i \,.
\label{GMeq5p6}
\end{equation}
In this expression the dependence on the scale $\mu$
appears exclusively through 
the argument of the $D$-dimensional coupling in $\widehat{\gamma}_K$ 
and $\widehat{\delta}_{\overline{\cal S}}$. Therefore \eqn{rencalS} is easily
integrated to give the corresponding formula for the reduced soft matrix
$\overline{{\cal S}}$,
\begin{align}
\begin{split}
\label{barS_ansatz}
\overline{{\cal S}}_{\dip}
\left(\rho_{i j}, \alpha_s,\epsilon\right) 
 &= \, \exp\Bigg\{
-\frac12 \int_0^{\mu^2} \frac{d\lambda^2}{\lambda^2} \, \Bigg[
 \, \frac12 \, \widehat{\delta}_{{\overline{\cal S}}}
( \alpha_s(\lambda^2,\epsilon) )  \,
\sum_{i = 1}^n  {\bf T}_i \cdot {\bf T}_i \, 
\\& \hskip1.5cm
- \frac18 \,
\widehat{\gamma}_K\left(\alpha_s(\lambda^2,\epsilon) \right) \,
\sum_{i = 1}^n \sum_{j \neq i} \,
\ln \rho_{ij} \, {\bf T}_i \cdot   {\bf T}_j\, 
\Bigg]\Bigg\}  \,\, .
\end{split}
\end{align}
\Eqn{GMeq5p6} satisfies the constraints~(\ref{constraints}) if and only
if the cusp anomalous dimension admits Casimir scaling, namely
$\gamma_K^{(i)}(\alpha_s) = C_i \, \widehat{\gamma}_K(\alpha_s)$, 
with $\widehat{\gamma}_K$ independent of the color representation 
of parton $i$.  In this paper we shall assume that this is the case,
postponing to future work the analysis of how higher-order Casimir 
contributions to $\gamma_K$ would affect the soft anomalous 
dimension matrix (the starting point for such an analysis is eq. (5.5) of 
Ref.~\cite{Gardi:2009qi}).

\begin{figure}[htb]
\begin{center}
\includegraphics[angle=0,width=5.0cm]{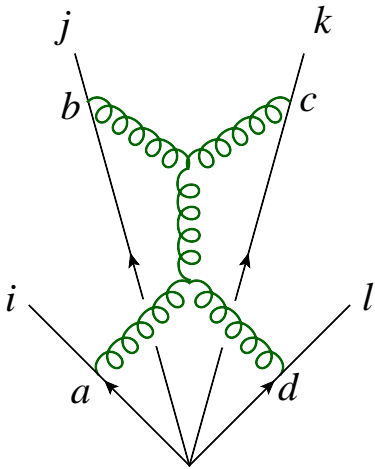}   
\caption{A purely gluonic diagram connecting the four hard partons
labeled $i,j,k,l$, which may contribute to the soft anomalous dimension
matrix at three loops. It correlates the colors of the four partons via
the operator
${\bf T}_i^{a} {\bf T}_j^{b} {\bf T}_k^{c} {\bf T}_l^{d}\ f^{ade} f^{cbe}$.
\label{4Elabeled}}
\end{center}
\end{figure}

Even under the assumption of Casimir scaling for $\gamma_K^{(i)}$,
\eqn{barS_ansatz} may not be the full result for $\overline{{\cal S}}$,
because $\Gamma^{{\overline{\cal S}}}$ may receive additional
corrections $\Delta^{{\overline{\cal S}}}$ going beyond the
sum-over-dipoles ansatz. In this case the full anomalous dimension can
be written as a sum,
\begin{equation}
\label{Gamma_barS}
\Gamma^{{\overline{\cal S}}}\left(\rho_{i j}, \alpha_s \right)
= \Gamma_{\dip}^{{\overline{\cal S}}}\left(\rho_{i j}, \alpha_s \right)
\,+\, \Delta^{{\overline{\cal S}}} \left(\rho_{i j}, \alpha_s \right)\,.
\end{equation}
Here $\Delta^{{\overline{\cal S}}}$ is a matrix in color space, which is
constrained to satisfy the homogeneous differential equation
\begin{equation}
\label{Delta_oureq_reformulated}
\sum_{j \neq i} \frac{\partial}{\partial \ln\rho_{ij}} 
\Delta^{{\overline{\cal S}}} \left( 
\rho_{i j}, \alpha_s \right) =  0 \, \qquad \forall i \,.
\end{equation}
This equation is solved by any function of
conformally invariant cross ratios of the form
\begin{equation}
\label{rhoijkl}
 \rho_{ijkl} \equiv \frac{\beta_i \cdot \beta_j \ \beta_k 
 \cdot \beta_l}{\beta_i \cdot \beta_k \ \beta_j \cdot \beta_l} \, ,
\end{equation}
which are related to the kinematic variables $\rho_{ij}$ in \eqn{rhoij}, and
to the momenta $p_i$, by
\begin{equation}
\label{rhoijkl_mod}
 \rho_{ijkl} 
  = \left(\frac{\rho_{i j} \, \rho_{k l}}{\rho_{i k} \, \rho_{j l}} 
 \right)^{1/2} 
 = \frac{p_i \cdot p_j \ p_k \cdot p_l}
        {p_i \cdot p_k \ p_j \cdot p_l} \, 
=\left|\frac{p_i \cdot p_j \ p_k \cdot p_l}
        {p_i \cdot p_k \ p_j \cdot p_l} \right| 
{\rm e}^{-{\rm i}\pi(\lambda_{ij} + \lambda_{kl}
                  -  \lambda_{ik} - \lambda_{jl})}.
\end{equation}
Each leg that appears in $\rho_{ijkl}$ does so once in the 
numerator and once in the denominator, thus cancelling in
the combination of derivatives in \eqn{Delta_oureq_reformulated}.
Hence we define
\begin{align}
\Delta^{{\overline{\cal S}}} \left( 
\rho_{i j}, \alpha_s \right)
= \Delta \left( \rho_{i j k l}, \alpha_s \right) \,.
\end{align}
Any additional correction $\Delta$ must introduce new correlations
between at least four partons into the reduced soft function.  
Such additional corrections are known not to appear at two
loops~\cite{Aybat:2006wq,Aybat:2006mz}, as expected from the 
fact that two-loop webs can correlate at most three hard partons. By 
the same token, they cannot show up in matter-dependent diagrams 
at three loops, as verified explicitly in Ref.~\cite{Dixon:2009gx}. 
On the other hand, they might be generated at three loops by purely 
gluonic diagrams, such as the one shown in~\fig{4Elabeled}.


The main purpose of the present paper 
is to examine all available constraints on the soft anomalous 
dimension matrix, and check whether they are sufficient to 
rule out a non-vanishing $\Delta$ at three loops. 
We will show that, despite the powerful constraints available, corrections 
to the sum-over-dipoles formula may indeed appear at this order.
In the case of purely logarithmic functions of the cross ratios
$\rho_{i j k l}$, we find a unique solution to all the constraints.
Allowing also for the appearance of polylogarithms of a single variable,
there are at least two additional solutions.
%


\section{Minimal ansatz for the singularities of the
amplitude~\label{sec:amplitude_ansatz}}

The factorization formula~(\ref{facamp_bar}) has the attractive
property that each of the singular factors is defined in a gauge-invariant
way.  It requires the introduction of the auxiliary vectors $n_i$,
which have been very useful~\cite{Gardi:2009qi} in revealing
the properties of the soft anomalous dimension.  At the end of the
day, however, the singularities of the amplitude ${\cal M}$
cannot depend on these auxiliary vectors, but only on the kinematic
invariants built out of external parton momenta. Indeed, as discussed
below, the cancellation of the dependence of the singular terms on the vectors
$n_i$ can be explicitly performed, and one can write the factorization 
of the amplitude in a more compact form:
\begin{equation}
\label{introducing_Z}
 {\cal M} \left(\frac{p_i}{\mu}, \alpha_s (\mu^2),  \epsilon \right)  =
 Z \left(\frac{p_i}{\mu_f}, \alpha_s(\mu_f^2),  \epsilon \right) \,\,
{\cal H} \left(\frac{p_i}{\mu}, \frac{\mu_f}{\mu}, \alpha_s(\mu^2),
\epsilon \right) \, ,
\end{equation}
as used in~Refs.~\cite{Becher:2009qa,Becher:2009cu}.
Here the (matrix-valued) $Z$ factor absorbs all the infrared
(soft and collinear) singularities, while the hard function ${\cal H}$ 
is finite as $\epsilon\rightarrow0$.  We distinguish between two
scales, the renormalization scale $\mu$, which is present in the
renormalized amplitude ${\cal M}$ on the left-hand side of
\eqn{introducing_Z}, and $\mu_f$, a factorization scale that is
introduced through the $Z$ factor. The function ${\cal H}$ (a vector 
in color space) plays the role of $H$ in the factorization 
formula~(\ref{facamp_bar}), but differs from it by being independent 
of the auxiliary vectors $n_i$.

Sudakov factorization implies that the $Z$ matrix is renormalized 
multiplicatively. We can then define the anomalous dimension matrix 
$\Gamma$, corresponding to $Z$, by
\begin{equation}
\label{Gamma_def}
\frac{d}{d \ln \mu_f} Z \left(\frac{p_i}{\mu_f}, \alpha_s(\mu_f^2),
\epsilon \right) \,
= \, - \,  
Z \left(\frac{p_i}{\mu_f}, \alpha_s(\mu_f^2), \epsilon \right)
 \, \Gamma \left(\frac{p_i}{\mu_f}, \alpha_s (\mu_f^2) \right) . 
\end{equation}
Note that the matrix $\Gamma$ is finite, but it can depend implicitly on
$\epsilon$ when evaluated as a function of the $D$-dimensional running 
coupling; it will then generate the infrared poles of $Z$, as usual, through
integration over the scale. 

The sum-over-dipoles ansatz for $\Gamma^{\overline{\cal S}}$, 
\eqn{GMeq5p6}, implies an analogous formula for $\Gamma$.
In order to see it, one may use the factorization formula~(\ref{facamp_bar}), 
substitute in \eqns{J_explicit}{barS_ansatz}, use color conservation,
$\sum_{j\neq i} {\bf T}_j=- {\bf T}_i$, and apply the identity
\begin{equation}
\label{kinematic_variables_combined}
\displaystyle{\underbrace{
\ln \left(\frac{(2p_i\cdot n_i)^2}{n_i^2}\right)}_{J_i}\,+\,
\underbrace{\ln \left(\frac{(2p_j\cdot n_j)^2}{n_j^2}\right)}_{J_j}}\,
\, + \,\underbrace{\ln \left(\frac{\left(\left|\beta_i\cdot\beta_j\right|\,\,
{\rm e}^{-{\rm i} \pi\lambda_{ij}}\right)^2}
{\displaystyle{\frac{2(\beta_i\cdot n_i)^2}{n_i^2}
\frac{2(\beta_j\cdot n_j)^2}{n_j^2}}}\right)}_{\overline{\cal S}}
= 2 \ln(2 \left| p_i\cdot p_j\right|\,{\rm e}^{-{\rm i} \pi\lambda_{ij}}) \, .
\end{equation}
Note also that the poles associated with
$\widehat{\delta}_{{\overline{\cal S}}}(\alpha_s)$
cancel out between the soft and jet functions. In this way, one arrives 
at the sum-over-dipoles ansatz for $\Gamma$,
\begin{align}
\label{Gamma_ansatz}
\begin{split}
\Gamma_{\dip} \left( \frac{p_i}{\lambda}, 
\alpha_s (\lambda^2) \right) = &- \frac14 \,
\widehat{\gamma}_K\left(\alpha_s(\lambda^2)  \right) 
\sum_{i =1}^n \sum_{j \neq i} \,
\ln\left(\frac{ \, 2 \, \left| p_i \cdot p_j\right| 
\,{\rm e}^{-{\rm i} \pi\lambda_{ij}}}
   {{\lambda^2}}\right) 
 {\bf T}_i \cdot   {\bf T}_j\, 
\\&+ \sum_{i=1}^n \, 
\gamma_{J_i} \left(\alpha_s(\lambda^2) \right) \,.
\end{split}
\end{align}
The $Z$ matrix which solves \eqn{Gamma_def} can be written in 
terms of the sum-over-dipoles ansatz~(\ref{Gamma_ansatz}) as an 
exponential, in a form similar to \eqn{barS_ansatz}.  
However, $\overline{\cal S}_{\dip}$ has only simple poles in $\e$ in
the exponent,
while the integration of $\Gamma_{\dip}$ over the scale $\lambda$ of the 
$D$-dimensional running coupling will generate double (soft-collinear) 
poles within $Z$, inherited from the jet functions in \eqn{J_explicit},
because of the explicit dependence of $\Gamma_{\dip}$ on the logarithm
of the scale $\lambda$.

If a non-trivial correction $\Delta$ appears in the reduced soft 
function~(\ref{Gamma_barS}), then the full anomalous dimension is
\begin{align}
\label{Gamma}
\Gamma \left(\frac{p_i}{\lambda}, \alpha_s(\lambda^2) \right)
= \Gamma_{\dip} \left(\frac{p_i}{\lambda}, \alpha_s(\lambda^2) \right)
\,+\, \Delta \left(\rho_{i j k l},\alpha_s(\lambda^2) \right)\,.
\end{align}
In terms of this function the solution of \eqn{Gamma_def} takes the form
\begin{equation}
\label{Z}
Z \left( \frac{p_i}{\mu_f}, \alpha_s(\mu_f^2), \epsilon \right)  \, = \,
\, {\rm P} \exp\Bigg\{
-\frac12 \int_0^{\mu_f^2}
\frac{d \lambda^2}{\lambda^2} \, \Gamma \left(\frac{p_i}{\lambda},
\alpha_s \left( \lambda^2, \epsilon \right) \right)
\Bigg\}  \,,
\end{equation}
where ${\rm P}$ stands for path-ordering: the order of the color
matrices after expanding the exponential coincides with the ordering in
the scale $\lambda$.  We emphasize that path ordering is only
necessary in \eqn{Z} if $\Delta\neq 0$ and
$[\Delta,\Gamma_{\dip}]\neq 0$.  Indeed, the ansatz~(\ref{Gamma_ansatz})
has the property that the scale-dependence associated with
non-trivial color operators appears through an overall factor, 
$\widehat\gamma_K(\alpha_s(\lambda^2))$, so
that color matrices $\Gamma$ corresponding to different scales are
proportional to each other, and obviously commute.  This is no longer
true for a generic $\Delta\neq 0$, starting at a certain loop order $l$. 
In this case \eqn{Gamma} would generically be a sum of two non-commuting
matrices, each of them having its own dependence on the coupling and thus 
on the scale $\lambda$.  Considering two scales $\lambda_1$ and 
$\lambda_2$, we would then have $[\Gamma(\lambda_1),
\Gamma(\lambda_2)] \neq 0$, and the order of the matrices
in the expansion of \eqn{Z} would be dictated by the ordering of the
scales. It should be noted, though, that the first loop order in $Z$ that
would be affected is order $l+1$, because $\Gamma$ starts at one loop,
so that
\begin{equation}
\label{noncommutativity}
\left[ \Gamma \left( \lambda_1 \right), 
        \Gamma \left( \lambda_2 \right)
\right] 
         \sim  
\left[ \Gamma^{(1)} \left( \lambda_1 \right), 
\Delta^{(l)} \left( \lambda_2 \right) \right]
= {\cal O} (\alpha_s^{l+1}) \, .
\end{equation}
The issue of ordering can thus be safely neglected at three loops, the first
order at which a non-vanishing $\Delta$ can arise.


\section{The splitting-amplitude constraint~\label{sec:SA}}

Let us now consider the limit where two of the hard partons in the
amplitude become collinear. Following Ref.~\cite{Becher:2009qa},
we shall see that this limit provides an additional constraint on the
structure of $\Delta$.  The way we use this constraint in the next
section will go beyond what was done in Ref.~\cite{Becher:2009qa}; we will
find explicit solutions satisfying the constraint (as well as other
consistency conditions discussed in the next section).

The Sudakov factorization described by \eqn{facamp_bar}, and subsequently the
singularity structure encoded in $Z$ in \eqn{introducing_Z}, apply
to scattering amplitudes at fixed angles. All the invariants 
$p_i\cdot p_j$ are taken to be of the same order, much larger than 
the confinement scale $\Lambda^2$. The limit in which
two of the hard partons are taken collinear, {\it e.g.} $p_1\cdot p_2\to 0$,
is a singular limit, which we are now about to explore. 
In this limit, $p_1 \to zP$ and $p_2 \to (1-z)P$, where the longitudinal
momentum fraction $z$ obeys $0<z<1$ (for time-like splitting). We will see,
following Ref.~\cite{Becher:2009qa}, that there is a
relation between the singularities that are associated with the
splitting --- the replacement of one parton by two collinear partons
--- and the singularities encoded in $Z$ in \eqn{introducing_Z}.

It is useful for our derivation to make a clear distinction between the two
scales $\mu_f$ and $\mu$ introduced in \eqn{introducing_Z}. Let us
first define the splitting amplitude, which relates the
dimensionally-regularized amplitude for the scattering of $n - 1$ 
partons to the one for $n$ partons, two of which are taken collinear.
We may write
\begin{equation} 
\label{Sp_M}
{\cal M}_n \left(p_1, p_2, p_j; \mu, \epsilon \right)
\ \iscol{1}{2} \ \, {\bf Sp}\left( p_1, p_2; \mu, \epsilon \right) \,
{\cal M}_{n-1} \left( P, p_j; \mu, \epsilon \right) \,.
\end{equation}
Here the two hard partons that become collinear are denoted by
$p_1$ and $p_2$, and all the other momenta by $p_j$, with $j = 3, 4, 
\ldots, n$.  We have slightly modified our notation for simplicity: 
the number of colored partons involved in the scattering is indicated
explicitly; the dependence of each factor on the running coupling
is understood; finally, the matrix elements have dimensionful arguments
(while in fact they depend on dimensionless ratios, as indicated in 
the previous sections). 
The splitting described by \eqn{Sp_M} preserves 
the total momentum $p_1 + p_2 = P$ and the total color charge $
{\bf T}_1 + {\bf T}_2 = {\bf T}$. We assume \eqn{Sp_M} to be valid
in the collinear limit, up to corrections that must be finite as 
$P^2 = 2 p_1 \cdot p_2 \to 0$.

The splitting 
matrix ${\bf Sp}$ encodes all singular contribution to the
amplitude ${\cal M}_n$ arising from the limit $P^2 \to 0$, and, crucially,
it must depend only on the quantum numbers of the splitting partons. The
matrix element ${\cal M}_{n - 1}$, in contrast, is evaluated at $P^2 = 0$,
and therefore it obeys Sudakov factorization, \eqn{introducing_Z}, as applied 
to an $(n-1)$-parton amplitude.
The operator ${\bf Sp}$ is designed to relate 
color matrices defined in the $n$-parton color space to those defined in
the $(n-1)$-parton space: it multiplies on its left the former and on
its right the latter. Thus, the initial definition of ${\bf Sp}$ is not
diagonal. Upon substituting ${\bf T} = {\bf T}_1 + {\bf T}_2$, however,
one can use the $n$-parton color space only.  In this space ${\bf Sp}$ 
is diagonal; all of its dependence on ${\bf T}_1$ and ${\bf T}_2$ can 
be expressed in terms of the quadratic Casimirs, using $2 \, {\bf T}_1 \cdot 
{\bf T}_2 = {\bf T}^2 - {\bf T}_1^2 - {\bf T}_2^2$.

Because the fixed-angle factorization theorem in \eqn{facamp_bar} breaks 
down in the collinear limit, $p_1 \cdot p_2 \to 0$, we expect that some 
of the singularities captured by the splitting matrix ${\bf Sp}$ will arise
from the hard functions ${\cal H}$. Specifically, if the $Z$ factor in
\eqn{introducing_Z} is defined in a minimal scheme, ${\cal H}$ will 
contain all terms in ${\cal M}_n$ with logarithmic singularities in 
$p_1 \cdot p_2$ associated with non-negative powers of $\epsilon$.
We then define ${\bf Sp}_{\cal H}$, in analogy with \eqn{Sp_M}, by
the collinear behavior of the hard functions,
\begin{equation} 
\label{Sp_H}
{\cal H}_n \left(p_1, p_2, p_j; \mu, \mu_f, \epsilon \right)
\ \iscol{1}{2} \ \, {\bf Sp}_{\cal H}(p_1, p_2; \mu, \mu_f, \epsilon) \, 
{\cal H}_{n - 1} \left(P, p_j;\mu, \mu_f, \epsilon \right) \, ,
\end{equation}
where all factors are finite as $\epsilon \to 0$. As was the case for 
\eqn{Sp_M}, \eqn{Sp_H} is valid up to corrections that remain finite in 
the limit $P^2 \to 0$. Singularities in that limit are all contained in the 
splitting matrix ${\bf Sp}_{\cal H}$, while the function ${\cal H}_{n - 1}$ 
is evaluated at $P^2 = 0$. 

Next, recall the definition of the $Z$ factors in \eqn{introducing_Z}
for both the $n$- and $(n - 1)$-parton amplitudes. In the present notation, 
they read
\begin{align}
\label{introducing_Z_n}
 {\cal M}_n \left(p_1, p_2, p_j; \mu, \epsilon \right)  &=
 Z_n \left(p_1, p_2, p_j; \mu_f, \epsilon \right) \,\,
{\cal H}_n \left(p_1, p_2, p_j; \mu, \mu_f, \epsilon \right) \,,
\\
\label{introducing_Z_n-1}
 {\cal M}_{n - 1} \left(P, p_j; \mu, \epsilon \right)  &=
 Z_{n - 1} \left(P, p_j; \mu_f, \epsilon \right) \,\,
{\cal H}_{n-1} \left(P, p_j; \mu, \mu_f, \epsilon \right) \,.
\end{align}
Substituting \eqn{introducing_Z_n-1} into \eqn{Sp_M} yields
\begin{equation}
{\cal M}_n \left(p_1, p_2, p_j; \mu, \epsilon \right)
\ \iscol{1}{2} \ \, {\bf Sp} (p_1, p_2; \mu, \epsilon) \,
 Z_{n - 1} \left( P, p_j; \mu_f, \epsilon \right) \,\,
{\cal H}_{n - 1} \left(P, p_j; \mu, \mu_f, \epsilon \right) \,.
\end{equation}
On the other hand, substituting \eqn{Sp_H}  into \eqn{introducing_Z_n} 
we get
\begin{equation}
 {\cal M}_n \left(p_1, p_2, p_j; \mu, \epsilon \right) 
\, \iscol{1}{2} \ Z_n \left(p_1, p_2, p_j; \mu_f, \epsilon \right) 
\,{\bf Sp}_{\cal H}(p_1, p_2; \mu, \mu_f, \epsilon) \, 
{\cal H}_{n-1} \left(P, p_j; \mu, \mu_f, \epsilon \right) \,.
\end{equation}
Comparing these two equations we immediately deduce the relation 
between the full splitting matrix ${\bf Sp}$, which is infrared divergent,
and its infrared-finite counterpart ${\bf Sp}_{\cal H}$,
\begin{align}
\label{Sp_Z_relation}
{\bf Sp}_{\cal H}(p_1, p_2; \mu, \mu_f, \epsilon) \, = \,
Z^{-1}_n \left(p_1, p_2, p_j; \mu_f, \epsilon \right)
\, {\bf Sp}(p_1, p_2; \mu, \epsilon) \,
Z_{n - 1} \left(P, p_j; \mu_f, \epsilon \right) \,,
\end{align}
where $Z_n$ is understood to be evaluated in the collinear limit.
This equation ({\it cf.}~eq.~(55) in Ref.~\cite{Becher:2009qa}) is a
non-trivial constraint on both $Z$ and the splitting amplitude
${\bf Sp}$, given that the left-hand side must be finite as $\epsilon 
\to 0$, and that the splitting amplitude depends only on the momenta 
and color variables of the splitting partons --- not on other hard partons 
involved in the scattering process.

To formulate these constraints, we take a logarithmic derivative 
of \eqn{Sp_Z_relation}, using the definition of $\Gamma_n$ and 
$\Gamma_{n - 1}$ according to \eqn{Gamma_def}. Using the fact
that ${\bf Sp}(p_1, p_2; \mu, \epsilon)$ does not depend on $\mu_f$,
it is straightforward to show that
\begin{align}
\label{Gamma_Sp_der}
\begin{split}
\frac{d}{d \ln \mu_f} 
\, {\bf Sp}_{\cal H} (p_1, p_2; \mu, \mu_f, \epsilon)\, = \,
 \, &\Gamma_n\left(p_1, p_2, p_j; \mu_f\right)   
\,  {\bf Sp}_{\cal H}(p_1, p_2; \mu, \mu_f, \epsilon) \\ &- 
\, {\bf Sp}_{\cal H}(p_1, p_2; \mu, \mu_f, \epsilon) \,  
\Gamma_{n - 1} \left(P, p_j; \mu_f \right) \, ,
\end{split}
\end{align}
where, as above, $(n - 1)$-parton matrices are evaluated in collinear
kinematics ($P^2 = 0$), and corrections are finite in the collinear
limit. Note that all the functions entering (\ref{Gamma_Sp_der}) are
finite for $\epsilon\to 0$. Note also that we have adapted the $\Gamma$
matrices to our current notation with dimensionful arguments; as before,
the matrices involved acquire implicit $\epsilon$ dependence when
evaluated as functions of the $D$-dimensional coupling.
 
Upon using the identification ${\bf T} = {\bf T}_1 + {\bf T}_2$, the
matrix $\Gamma_{n - 1}$ can be promoted to operate on the $n$-parton color
space.  Once one does this, one immediately recognizes that the
splitting matrix
${\bf Sp}$ commutes with the $\Gamma$ matrices, as an immediate
consequence of the fact that it can only depend on the color degrees of
freedom of the partons involved in the splitting, {\it i.e.} ${\bf T}_1$,
${\bf T}_2$ and ${\bf T} = {\bf T}_1 + {\bf T}_2$, and it is therefore
color diagonal.  Therefore, we can rewrite \eqn{Gamma_Sp} as an evolution
equation for the splitting amplitude:
\begin{align}
\label{SpHdiffeq}
\frac{d}{d\ln\mu_f} \, {\bf Sp}_{\cal H} (p_1, p_2; \mu, \mu_f, \epsilon) \,
= \, \Gamma_{\bf Sp} (p_1, p_2; \mu_f) \, \,
{\bf Sp}_{\cal H}(p_1, p_2; \mu, \mu_f, \epsilon) \, ,
\end{align}
where we defined
\begin{align}
\label{Gamma_Sp}
\Gamma_{\bf Sp}(p_1, p_2; \mu_f)
\equiv \Gamma_n \left(p_1, p_2, p_j; \mu_f \right) -
\Gamma_{n - 1} \left(P, p_j; \mu_f \right) \, .
\end{align}

We may now solve \eqn{SpHdiffeq} for the $\mu_f$ dependence of 
${\bf Sp}_{\cal H}$, with the result
\begin{align}
\label{solu}
{\bf Sp}_{\cal H} (p_1, p_2; \mu, \mu_f, \epsilon) \, = \, 
{\bf Sp}_{\cal H}^{(0)} (p_1, p_2; \mu, \epsilon) \, \exp \left[
\frac12 \int_{\mu^2}^{\mu_f^2} \frac{d \lambda^2}{\lambda^2} \, 
\Gamma_{\bf Sp} (p_1, p_2; \lambda)
\right] \, .
\end{align}
The initial condition for evolution
\begin{equation}
\label{inicon}
{\bf Sp}_{\cal H}^{(0)} (p_1, p_2; \mu, \epsilon) \, = \, 
{\bf Sp}_{\cal H} (p_1, p_2; \mu, \mu_f = \mu, \epsilon)
\end{equation}
will, in general, still be singular as $p_1 \cdot p_2 \to 0$, although
it is finite as $\epsilon \to 0$. We may, in any case, use \eqn{solu}
by matching the $\mu$-dependence in \eqn{Sp_Z_relation}, which yields
an expression for the full splitting function ${\bf Sp}$. We find
\begin{align}
\label{solusp}
{\bf Sp} (p_1,p_2; \mu, \epsilon) \, = \, 
{\bf Sp}_{\cal H}^{(0)} (p_1, p_2; \mu, \epsilon)  \, \exp \left[
- \frac12 \int_{0}^{\mu^2} \frac{d \lambda^2}{\lambda^2} \, 
\Gamma_{\bf Sp} (p_1, p_2; \lambda)
\right] \, .
\end{align}
While collinear singularities accompanied by non-negative powers 
of $\epsilon$ are still present in the initial condition, all poles in
$\epsilon$ in the full splitting matrix arise from the integration 
over the scale of the $D$-dimensional running coupling in the exponent
of \eqn{solusp}. 

The restricted kinematic dependence of $\Gamma_{\bf Sp}$,
which generates the poles in the splitting function ${\bf Sp}$,
is sufficient to provide nontrivial constraints
on the matrix $\Delta$, as we will now see. Indeed,
substituting \eqn{Gamma} into \eqn{Gamma_Sp} we obtain
\begin{align}
\label{Gamma_Sp_explicit}
\Gamma_{\bf Sp}(p_1, p_2; \lambda) \, 
= \, \Gamma_{{\bf Sp}, \, {\dip}} (p_1, p_2; \lambda)
+ \Delta_n \left(\rho_{i j k l}; \lambda \right) 
- \Delta_{n - 1} \left(\rho_{i j k l}; \lambda \right) \, ,
\end{align}
where
\begin{align}
\label{Gamma_Sp_ansatz_explicit}
\begin{split}
\Gamma_{{\bf Sp}, \,{\dip}}(p_1, p_2; \lambda)
= & - \frac12 \, \widehat{\gamma}_K \left(\alpha_s (\lambda^2)
\right) \Bigg[
\ln \left(\frac{2\left| p_1 \cdot p_2\right| 
\, {\rm e}^{-{\rm i} \pi\lambda_{12}}}
{{\lambda^2}}\right) \,  {\bf T}_1 \cdot   {\bf T}_2\, \\
& -  {\bf T}_1 \cdot\left({\bf T}_1 + {\bf T}_2 \right) \ln z
- {\bf T}_2 \cdot\left({\bf T}_1 + {\bf T}_2 \right) \ln(1 - z) \Bigg]
\\  & +  \, \gamma_{J_1} \left(\alpha_s(\lambda^2) \right) 
+ \, \gamma_{J_2} \left(\alpha_s(\lambda^2) \right) 
- \, \gamma_{J_P} \left(\alpha_s(\lambda^2) \right) \, .
\end{split}
\end{align}
\Eqn{Gamma_Sp_ansatz_explicit}
is the result of substituting the sum-over-dipoles
ansatz~(\ref{Gamma_ansatz}) for $\Gamma_n$ and $\Gamma_{n-1}$.
The terms in \eqn{Gamma_Sp_explicit} going beyond
\eqn{Gamma_Sp_ansatz_explicit} depend on conformally invariant cross
ratios in the $n$-parton and $(n-1)$-parton amplitudes, respectively. Their
difference should conspire to depend only on the kinematic variables
$p_1$ and $p_2$ and on the color variables ${\bf T}_1$ and
${\bf T}_2$. In this way \eqn{Gamma_Sp_explicit} provides a
non-trivial constraint on the structure of~$\Delta$, which we will
implement in \sect{CollimitSubsection}.


\section{Constraining corrections to the sum-over-dipoles
formula~\label{sec:corrections}}


\subsection{Functions of conformally invariant cross ratios} 

Our task here is to analyze potential contributions of the form
$\Delta\left(\rho_{i j k l}, \alpha_s \right)$ to the soft
singularities of any $n$-leg amplitude. Our starting point is the fact
that these contributions must be written as functions
of conformally invariant cross ratios of the form~(\ref{rhoijkl}). 
Because we are dealing with renormalizable theories in four dimensions,
we do not expect $Z$ to contain power-law dependence on the kinematic
variables; instead the dependence should be ``slow'', {\it i.e.} logarithmic
in the arguments, through variables of the form
\begin{align}
\label{lnrho1234}
L_{ijkl}\ \equiv\ \ln\rho_{ijkl}
\ =\ \ln \left(\frac{p_i\cdot p_j \, \, p_k \cdot p_l}{p_i
\cdot p_k \, \, p_j \cdot p_l} \right) \,.
\end{align}
Eventually, at high enough order, dependence on $\rho_{ijkl}$
through polylogarithms and harmonic polylogarithms might arise.
We will not assume here that $\Delta$ is linear in the variables $L_{ijkl}$.  
We will allow logarithms of  different cross ratios to appear in a product, 
raised to various powers, and this will be a key to finding solutions 
consistent with the collinear limits. Subsequently, we will examine how
further solutions may arise if polylogarithmic dependence is allowed.

A further motivation to consider a general logarithmic dependence 
through the variables in \eqn{lnrho1234} is provided by the collinear limits,
which can take certain cross ratios $\rho_{ijkl}$ to 0, 1, or $\infty$, 
corresponding to physical limits where logarithmic divergences in $\Delta$
will be possible. Other values of the cross ratios, on the other hand, 
should not cause (unphysical) singularities in $\Delta(\rho_{ijkl})$.
This fact limits the acceptable functional forms.  For example, in specifying 
a logarithmic functional dependence to be through \eqn{lnrho1234}, we 
explicitly exclude the form $\ln (c + \rho_{ijkl})$ for general\footnote{In
\sect{sec:poly} we will briefly consider the possibility of including a 
dependence of the form $\ln (1 - \rho)$.} constant $c$. Such a shift 
in the argument of the logarithm would generate unphysical
singularities at $\rho_{ijkl} = - c$, and would also lead to
complicated symmetry properties under parton exchange, which would 
make it difficult to accomodate Bose symmetry. We will thus focus our 
initial analysis on kinematic dependence through the variables $L_{ijkl}$.
Although it seems less natural, polylogarithmic dependence on
$\rho_{ijkl}$ cannot be altogether ruled out, and will be considered in 
the context of the three-loop analysis in \sect{sec:3loop}.

The fact that the variables~(\ref{lnrho1234}) involve the momenta of 
four partons, points to their origin in webs that connect (at least) four 
of the hard partons in the process, exemplified by \fig{4Elabeled}.
The appearance of such terms in the exponent, as a correction to 
the sum-over-dipoles formula, implies, through the non-Abelian 
exponentiation theorem, that they cannot be reduced to sums 
of independent webs connecting just two or three partons, 
neither diagrammatically nor algebraically. Indeed, for amplitudes
composed of just three partons the sum-over-dipoles formula is 
exact~\cite{Gardi:2009qi}. Similarly, because two-loop webs can connect 
at most three different partons, conformally invariant cross ratios cannot 
be formed. Consequently, at two loops there are no corrections to the
sum-over-dipoles formula, independently of the number of legs.  Thus, the 
first non-trivial corrections can appear at three loops, and if they appear, 
they are directly related to webs that connect four partons. For the 
remainder of this section, therefore, we will focus on corrections to the
sum-over-dipoles formula that arise from webs connecting precisely four
partons, although other partons or colorless particles can be present
in the full amplitude. Our conclusions are fully general at three loops,
as discussed in \sect{sec:n-leg}, because at that order no web can 
connect more than four partons.

We begin by observing that, independently of the loop order at which
four-parton corrections appear, their color factor must involve at least
one color generator corresponding to each of the four partons
involved. For example, the simplest structure a term in $\Delta$ can 
have in color space is
\begin{equation}
\label{color}
\Delta_4 (\rho_{ijkl}) \, = \, h^{abcd} \, {\bf T}_i^{a} \, {\bf T}_j^{b} \,
{\bf T}_k^{c} \, {\bf T}_l^{d} \, \Delta^{\kin}_4 (\rho_{ijkl}) \, ,
\end{equation} 
where $h^{abcd}$ is some color tensor built out of structure constants
corresponding to the internal vertices in the web 
that connects the four partons $(i,j,k,l)$ to each other. Note that
$h^{abcd}$ may receive contributions from several different webs at a
given order, and furthermore, for a given $h^{abcd}$, the kinematic 
coefficient $\Delta^{\kin}_4 (\rho_{ijkl})$ can receive corrections from 
higher-order webs. In what follows, we will not display the dependence
on the coupling of the kinematic factors, because it does not affect our 
arguments. As we will see, symmetry arguments will, in general, force
us to consider sums of terms of the form~(\ref{color}), with different
color tensors $h^{abcd}$ associated with different kinematic factors.

More generally, at sufficiently high orders, there can be other 
types of contributions in which each Wilson line in the soft function is 
attached to more than one gluon, and hence to more than one index in 
a color tensor. Such corrections will be sums of terms of the form
\begin{align}
\begin{split}
\label{color_}
\Delta_4 (\rho_{ijkl}) \, = \, &
{\Delta}^{\kin}_4 (\rho_{ijkl}) \,\, h^{a_1, \ldots, a_{m_1},
b_1, \ldots, b_{m_2},
c_1, \ldots, c_{m_3},
d_1, \ldots, d_{m_4} } \\ & 
({\bf T}_i^{a_1}{\bf T}_i^{a_2}\ldots {\bf T}_i^{a_{m_1}}
{\bf T}_j^{b_1}{\bf T}_j^{b_2}\ldots {\bf T}_j^{b_{m_2}}
{\bf T}_k^{c_1}{\bf T}_k^{c_2}\ldots {\bf T}_k^{c_{m_3}}
{\bf T}_l^{d_1}{\bf T}_l^{d_2}\ldots {\bf T}_l^{d_{m_1}})_{+}\,,
\end{split}
\end{align} 
where $()_{+}$ indicates symmetrization with respect to all the
indices corresponding to a given parton.  Note that generators
carrying indices of different partons commute, while the antisymmetric
components have been excluded from \eqn{color_}, because they reduce,
via the commutation relation $[{\bf T}_i^a \,,\, {\bf T}_i^b ] = i 
f^{abc} {\bf T}_i^c$, to shorter strings\footnote{One can 
make a stronger statement for Wilson lines in the fundamental representation 
of the gauge group.  In that case the symmetric combination in 
\eqn{color_} can also be further reduced, using the identity $\{{\bf t}_a ,
{\bf t}_b \} = \frac{1}{N_c} \delta_{a b} + d_{abc} {\bf t}_c$, so that
the generic correction in \eqn{color_} turns into a combination of terms
of the form~(\ref{color}). We are not aware of generalizations of this 
possibility to arbitrary representations.}. In the following subsections, we 
will focus on (combinations of) color structures of the form~(\ref{color}),
and we will not consider further the more general case of \eqn{color_},
which, in any case, can only arise at or beyond four loops.


\subsection{Bose symmetry}

The Wilson lines defining the reduced soft matrix are effectively
scalars, as the spin-dependent parts have been stripped off and absorbed
in the jet functions.  Consequently, the matrices $\Gamma$ and $\Delta$
should admit Bose symmetry and be invariant under the
exchange of any pair of hard partons.  Because $\Delta$
depends on color and kinematic variables, this symmetry
implies correlation between color and kinematics. In particular,
considering a term of the form~(\ref{color}), the symmetry properties 
of $h^{abcd}$ under permutations of the indices $a$, $b$, $c$ and $d$ 
must be mirrored in the symmetry properties of the kinematic factor
${\Delta}^{\kin}_4 (\rho_{ijkl})$ under permutations of the
corresponding momenta $p_i$, $p_j$, $p_k$ and $p_l$. The
requirement of Bose symmetry will lead us to express $\Delta$ as a 
sum of terms, each having color and kinematic factors with a definite
symmetry under some (or all) permutations.

Because we are considering corrections arising from four-parton webs,
we need to analyze the symmetry properties under particle exchanges 
of the ratios $\rho_{ijkl}$ that can be constructed with four partons.
There are 24 different cross ratios of this type, corresponding to the 
number of elements of the permutation group acting on four objects, 
$S_4$.  However, a $Z_2 \times Z_2$ subgroup of $S_4$ leaves 
each $\rho_{ijkl}$ (and hence each $L_{ijkl}$) invariant. Indeed,
one readily verifies that
\begin{equation}
\label{invrho}
\rho_{ijkl} = \rho_{jilk} = \rho_{klij} = \rho_{lkji} \,.
\end{equation}
The subgroup $Z_2 \times Z_2$ is an invariant subgroup of $S_4$. 
Thus, we may use it to fix one of the indices, say $i$, in $\rho_{ijkl}$.
This leaves six cross ratios, transforming under the permutation group of 
three objects, $S_3 \simeq S_4/(Z_2\times Z_2)$.


\begin{figure}[htb]
\begin{center}
\includegraphics[angle=0,width=6cm]{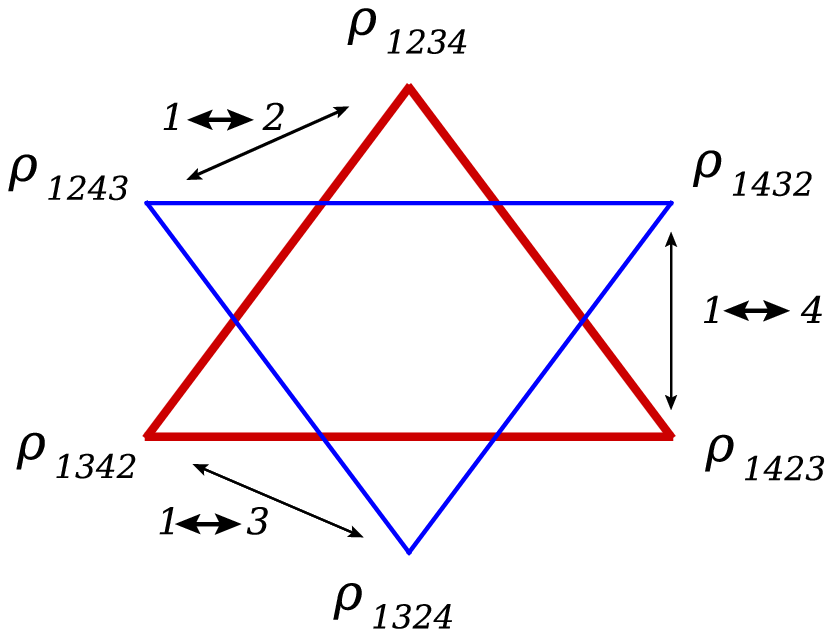}  \hspace*{20pt}  
\includegraphics[angle=0,width=8cm]{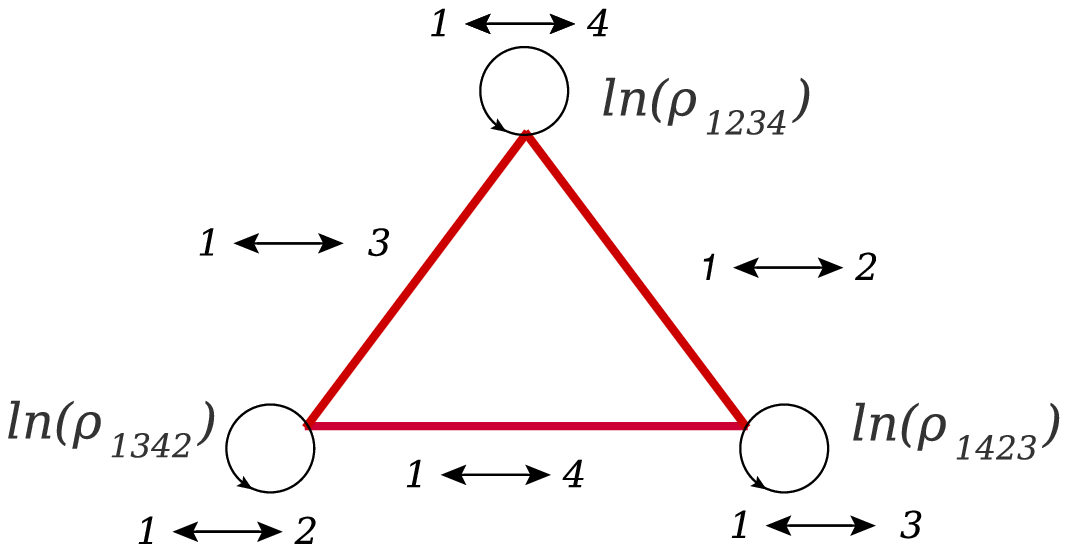}    
\caption{Symmetry properties of conformally invariant cross ratios. 
Each of the two triangles in the left-hand figure connects three cross
ratios. The cross ratios associated with the two triangles are related 
by inversion, and one moves between the two triangles with odd 
permutations of momentum labels. The right-hand figure shows the 
resulting antisymmetry of the logarithms of the three conformally 
invariant cross ratios under permutations.  The three variables 
$L_{1234}$, $L_{1423}$ and $L_{1342}$ transform into one 
another under permutations, up to an overall minus sign. For example, 
under the permutation $1\lr2$, we have
$L_{1234} \to - L_{1423}$,
$L_{1423} \to - L_{1234}$, and
$L_{1342} \to - L_{1342}$. 
\label{symm}}
\end{center}
\end{figure}


The permutation properties of the remaining six cross ratios are displayed
graphically in \fig{symm}, where we made the identifications $\{i,j,k,l\} 
\to \{1,2,3,4\}$ for simplicity. The analysis can be further simplified by 
noting that odd permutations in $S_4$ merely invert $\rho_{ijkl}$, so 
that, for example,
\begin{equation}
\label{odd_symmetry_of_rho}
\rho_{ijkl} \, = \, \frac{1}{\rho_{ikjl}} \qquad  \longrightarrow \qquad
L_{ijkl} \, = \, - \, L_{ikjl} \, .
\end{equation}
This inversion corresponds to moving across to the
diametrically opposite point in the left-hand plot in \fig{symm}.
We conclude that there are only three different cross ratios 
(corresponding to the cyclic permutations of $\{j,k,l\}$ associated with
$S_3/Z_2 \simeq Z_3$), namely $\rho_{ijkl}$, $\rho_{iljk}$ and 
$\rho_{iklj}$. They correspond to triangles in \fig{symm}.
Finally, the logarithms of the three cross ratios are linearly 
dependent, summing to zero:
\begin{equation}
\label{three_rho_relation}
L_{ijkl} + L_{iljk} + L_{iklj} \,=\,0\,.
\end{equation}

These symmetry properties lead us to consider for $\Delta^{\kin}_4$ in 
\eqn{color} the general form
\begin{align}
\label{kin_4legs}
\Delta_4^{\kin}(\rho_{ijkl})
= ( L_{1234} )^{h_1} \, ( L_{1423} )^{h_2}
\, ( L_{1342} )^{h_3} \, ,
\end{align}
where we have adopted  the labeling of hard partons by $\{1,2,3,4\}$
as in \fig{symm}. Here the $h_i$ are non-negative integers, and 
\eqn{three_rho_relation} has not yet been taken into account. 
Our general strategy will be to construct linear combinations of the 
monomials in \eqn{kin_4legs} designed to match the symmetries of 
the available color tensors, $h^{abcd}$ in \eqn{color}. Such combinations 
can be constructed for general $h_i$. As we shall see, however, 
transcendentality constraints restrict the integers $h_i$ to be small at 
low loop orders. In the three-loop case, this will suffice to eliminate  
all solutions to the constraints, except for a single function.

We begin by noting that the antisymmetry of $L_{1234}$ under 
the permutation $1 \lr 4$ (or under $2 \lr 3$, see \fig{symm}) is mirrored 
by the antisymmetry of the color factor $h^{abcd} \, {\bf T}_1^{a} 
{\bf T}_2^{b} {\bf T}_3^{c} {\bf T}_4^{d}\,$ if the tensor 
$h^{abcd} = f^{ade} f^{cbe}$, where $f^{ade}$ are the usual, 
fully antisymmetric ${\rm SU}(N_c)$ structure constants. The same 
is obviously true for any odd power of $L_{1234}$, while in the
case of even powers, an appropriate type of color tensor is
$h^{abcd}=d^{ade}d^{cbe}$, where $d^{ade}$ are the fully symmetric
${\rm SU} (N_c)$ tensors. Fig.~\ref{symm}, however, shows that under 
other permutations, the different cross ratios transform into one another. 
Therefore, if we are to write a function with a definite symmetry under all 
permutations, it must be a function of all three variables. Specifically, 
in order for a term of the form~(\ref{kin_4legs}) to have, by itself, a 
definite symmetry under all permutations, the powers $h_1$, $h_2$ and 
$h_3$, must all be equal. Alternatively, one can consider a linear combination 
of several terms of the form~(\ref{kin_4legs}), yielding together a function 
of the kinematic variables with definite symmetry. In this respect it is
useful to keep in mind that the sum of the three logarithms (all with a 
single power) is identically zero, by \eqn{three_rho_relation}.

Let us now construct the different structures that realize Bose
symmetry, by considering linear combinations of terms of the form 
of \eqn{color}, with $\Delta^{\kin}_4$ given by \eqn{kin_4legs}.
We consider first three examples, where the logarithms
$L_{ijkl}$ are raised to a single power $h$. As we will see, none of 
these examples will satisfy all the constraints; they are useful however
for illustrating the available structures.
\begin{itemize}
\item[a)]{} We first consider simply setting $h_1 = h_2 = h_3$ in 
\eqn{kin_4legs}, obtaining
\begin{equation}
\label{Delta_a}
\Delta_4 (\rho_{ijkl}) \, = \, h^{abcd} \, \, {\bf T}_1^{a} {\bf T}_2^{b} 
{\bf T}_3^{c} {\bf T}_4^{d} \,\,
\Big[ L_{1234} \,L_{1423}\,L_{1342}\Big]^{h}\,.
\end{equation} 
For odd $h$ the color tensor $h^{abcd}$ must be completely 
antisymmetric in the four indices, while for even $h$ it must be completely
symmetric. We anticipate that odd $h$ is ruled out, because completely 
antisymmetric four-index invariant tensors do not exist for simple Lie  
groups~\cite{deAzcarraga:1997ya}.  Furthermore, while symmetric 
tensors do exist, \eqn{Delta_a} is ruled out at three loops, because 
from \fig{4Elabeled} it is clear that only $h^{abcd} = f^{ade} f^{cbe}$ 
(or permutations thereof) can arise in Feynman diagrams at this order. 
\item[b)]{}
Our second example is
\begin{align}
\label{Delta_b}
\begin{split}
\Delta_4 (\rho_{ijkl}) \, = \,
\,{\bf T}_1^{a} {\bf T}_2^{b} {\bf T}_3^{c} {\bf T}_4^{d}\,
\,\Big[&f^{ade}f^{cbe} L_{1234}^{h}
+f^{cae}f^{dbe} L_{1423}^{h}
+f^{bae}f^{cde}  L_{1342}^{h}\Big] \,,
\end{split}
\end{align}
where $h$ must be odd. Alternatively, each $f^{ade}$ may be replaced 
by the fully symmetric ${\rm SU}(N_c)$ tensor $d^{ade}$, and then 
$h$ must be  even. In \eqn{Delta_b} each term has a definite symmetry 
only with respect to certain permutations, but the three terms transform 
into one another in such a way that their sum admits full Bose symmetry. 
We will see shortly that the structure in \eqn{Delta_b} does not satisfy 
the collinear constraints.
\item[c)]{}
Finally, one may consider the case where two of the three logarithms
appear together in a product, raised to some power $h$,
\begin{align}
\label{Delta_c}
\begin{split}
\Delta_4 (\rho_{ijkl}) &=
\,{\bf T}_1^{a} {\bf T}_2^{b} {\bf T}_3^{c} {\bf T}_4^{d}\,
\,\,\Big[ d^{abe}d^{cde} ( L_{1234}\, L_{1423} )^{h}\\
&\hskip2.6cm
+d^{dae}d^{cbe}        ( L_{1423}\, L_{1342} )^{h}
+d^{cae}d^{bde}         ( L_{1342}\, L_{1234} )^{h}
\Big]\,.
\end{split}
\end{align}
Once again, we observe that these color tensors cannot arise in 
three-loop webs. Furthermore, as we will see, \eqn{Delta_c}, 
at any loop order, fails to satisfy the collinear constraints.  
\end{itemize}

We are led to consider more general structures, using 
\eqns{color}{kin_4legs} with arbitrary integers $h_i$. As announced, 
we will satisfy Bose symmetry by constructing polynomial kinematical 
factors mimicking the symmetry of the available color tensors. One
may write for example
\begin{align}
\label{Delta_gen}
\begin{split}
&\Delta_4 (\rho_{ijkl}) \, = \, 
\, {\bf T}_1^{a} {\bf T}_2^{b} {\bf T}_3^{c} {\bf T}_4^{d} \\
&\hskip1cm
\times \hskip2.5mm \left[ \, f^{ade}f^{cbe} \,  L_{1234}^{h_1} \, \left(
L_{1423}^{h_2} \, L_{1342}^{h_3} \, - \, (-1)^{h_1 + h_2 + h_3} \, 
L_{1342}^{h_2} \, L_{1423}^{h_3} \right) \right. \\
&\hskip1.5cm
+ f^{cae}f^{dbe} \, L_{1423}^{h_1} \, \left(
L_{1342}^{h_2} \, L_{1234}^{h_3} \, - \, (-1)^{h_1 + h_2 + h_3} \, 
L_{1234}^{h_2} \, L_{1342}^{h_3} \right) \\
&\hskip1.5cm
+ \left. f^{bae}f^{cde} \, L_{1342}^{h_1} \, \left(
L_{1234}^{h_2} \, L_{1423}^{h_3} \, - \, (-1)^{h_1 + h_2 + h_3} \, 
L_{1423}^{h_2} \, L_{1234}^{h_3} \right) \,
\right] \, ,
\end{split}
\end{align}
where $h_1$, $h_2$ and $h_3$ can be any non-negative integers.
The first line is invariant, for example, under the permutation $1\lr4$ 
(when applied to both kinematics and color), the second line is invariant 
under $1\lr3$, and the third is invariant under $1\lr2$. The other exchange
symmetries are realized by the transformation of two lines
into one another.  For example, under $1\lr4$ the second line
transforms into the third and vice versa. In \eqn{Delta_gen} the
color and kinematic factors in each line are separately antisymmetric
under the corresponding permutation. Note that \eqn{Delta_b}
corresponds to the special case where $h_1$ in \eqn{Delta_gen} is
odd, while $h_2 = h_3 = 0$.

One can also construct an alternative Bose symmetrization 
using the symmetric combination,
\begin{align}
\label{Delta_gen_symm}
\begin{split}
&\Delta_4 (\rho_{ijkl}) \, = \, 
\, {\bf T}_1^{a} {\bf T}_2^{b} {\bf T}_3^{c} {\bf T}_4^{d} \\
&\hskip1cm
\times \hskip2.5mm \left[ \, d^{ade}d^{cbe} \,  L_{1234}^{h_1} \, \left(
L_{1423}^{h_2} \, L_{1342}^{h_3} \, + \, (-1)^{h_1 + h_2 + h_3} \, 
L_{1342}^{h_2} \, L_{1423}^{h_3} \right) \right. \\
&\hskip1.5cm
+ d^{cae}d^{dbe} \, L_{1423}^{h_1} \, \left(
L_{1342}^{h_2} \, L_{1234}^{h_3} \, + \, (-1)^{h_1 + h_2 + h_3} \, 
L_{1234}^{h_2} \, L_{1342}^{h_3} \right) \\
&\hskip1.5cm
+ \left. d^{bae}d^{cde} \, L_{1342}^{h_1} \, \left(
L_{1234}^{h_2} \, L_{1423}^{h_3} \, + \, (-1)^{h_1 + h_2 + h_3} \, 
L_{1423}^{h_2} \, L_{1234}^{h_3} \right) \,
\right] \,.
\end{split}
\end{align}
Note that \eqn{Delta_c} is reproduced by setting $h_1 = 0$ and 
$h_2 = h_3 = h$ in \eqn{Delta_gen_symm}.

Eqs.~(\ref{Delta_gen}) and (\ref{Delta_gen_symm}) both yield non-trivial 
functions for both even and odd powers $h_i$, with the following exceptions:
For even $h_1$, \eqn{Delta_gen} becomes identically zero if $h_2=h_3$; 
similarly, for odd $h_1$ \eqn{Delta_gen_symm} becomes identically zero 
if $h_2 = h_3$. It is interesting to note that \eqn{Delta_a} with odd $h$ 
cannot be obtained as a special case of \eqn{Delta_gen}.  Indeed, by choosing
$h_1 = h_2 = h_3$ one obtains the correct kinematic dependence, but then
the color structure factors out and vanishes by the Jacobi identity,
\begin{equation}
\label{Jacobi}
h^{abcd}=f^{ade}f^{cbe} 
+f^{cae}f^{dbe} 
+f^{bae}f^{cde} = 0 \, .
\end{equation}
In contrast, for even $h$, \eqn{Delta_a} can be obtained as a special 
case of \eqn{Delta_gen_symm}, setting
\begin{equation}
\label{f_dd}
h^{abcd}=d^{ade}d^{cbe} 
+d^{cae}d^{dbe} 
+d^{bae}d^{cde}\,,
\end{equation}
which is totally symmetric, as required. This is expected from the general
properties of symmetric and antisymmetric invariant tensors for simple Lie
algebras~\cite{deAzcarraga:1997ya}.

At any fixed number of loops $l$, the total power of the logarithms in 
\eqns{Delta_gen}{Delta_gen_symm},  $h_{\rm tot} 
\equiv h_1 + h_2 + h_3$, will play an important role. 
Indeed, $h_{\rm tot}$ 
is the degree of transcendentality of the function $\Delta_4$, as
defined in the Introduction, and it is
bounded from above by the maximal allowed transcendentality of the
anomalous dimension matrix at $l$ loops, as described in \sect{sec:maxtran}.
We expect then that at $l$ loops there will be a finite number of sets of 
integers $h_i$ satisfying the available constraints. The most general 
solution for the correction term $\Delta_4$ will then be given 
by a linear combination of symmetric and antisymmetric polynomials such
as those given in \eqns{Delta_gen}{Delta_gen_symm},
with all allowed choices of $h_i$.  

Such combinations include also contributions related to higher-order
Casimir operators. Indeed, summing 
over permutations of $\{h_1,h_2,h_3\}$ in the symmetric version of 
$\Delta_4$, \eqn{Delta_gen_symm}, one finds a completely symmetric 
kinematic factor, multiplying a color tensor which is directly related to the 
quartic Casimir operator (with a suitable choice of basis in the space of 
symmetric tensors over the Lie algebra~\cite{deAzcarraga:1997ya}),
\begin{align}
\label{Delta_gen_symm_alt}
\begin{split}
&\Delta_4 (\rho_{ijkl}) \, = \,
\, {\bf T}_1^{a} {\bf T}_2^{b} {\bf T}_3^{c} {\bf T}_4^{d}
\times \Big[ d^{ade}d^{cbe} + d^{cae}d^{dbe} + d^{bae}d^{cde}\Bigr]\\
&\hskip0.2cm \times
\Big[ 
L_{1234}^{h_1}\, L_{1423}^{h_2}\, L_{1342}^{h_3}
+ L_{1423}^{h_1}\, L_{1342}^{h_2}\, L_{1234}^{h_3}
+ L_{1342}^{h_1}\, L_{1234}^{h_2}\, L_{1423}^{h_3}\\
&\hskip0.4cm + (-1)^{h_1+h_2+h_3}\Big( 
L_{1234}^{h_1}\, L_{1342}^{h_2}\, L_{1423}^{h_3}
+ L_{1423}^{h_1}\, L_{1234}^{h_2}\, L_{1342}^{h_3}
+ L_{1342}^{h_1}\, L_{1423}^{h_2}\, L_{1234}^{h_3}\Big)
\Big] \,.
\end{split}
\end{align}
For even $h_{\rm tot} \equiv h_1 + h_2 + h_3$ this function is always 
non-trivial, while for odd $h_{\rm tot}$ it is only non-trivial if all three 
powers $h_i$ are different. We note once again that, due to the Jacobi 
identity~(\ref{Jacobi}), \eqn{Delta_gen_symm_alt} does not have an 
analog involving the antisymmetric structure constants.


\subsection{Maximal transcendentality}
\label{sec:maxtran}

Our next observation is that, at a given loop order, the total power of
the logarithms, $h_{\rm tot}$, cannot be arbitrarily high. It is well
known (although not proven mathematically)
that the maximal transcendentality $\tau_{\rm max}$ of the coefficient 
of the $1/\e^k$ pole in an $l$-loop amplitude (including $k=0$) is 
$\tau_{\rm max} = 2 l - k$.  If a function is purely logarithmic, this
value corresponds to $2l-k$ powers of logarithms.  In general, the space of 
possible transcendental functions is not fully characterized mathematically,
particularly for functions of multiple dimensionless arguments.
At the end of this subsection we give some examples of functions of definite
transcendental weight, which appear in scattering amplitudes for massless
particles, and which therefore might be considered candidates from
which to build solutions for $\Delta$.

Because $\Gamma$, $\Gamma_{\bf Sp}$ and $\Delta$ are associated with
the $1/\e$ single pole, their maximal
transcendentality is $\tau_{\rm max} = 2 l - 1$. For $\NeqFour$ 
super-Yang-Mills theory, in every known instance the terms arising in this 
way are purely of this maximal transcendentality: there are no
terms of lower transcendentality.  This property
is relevant also for non-supersymmetric massless gauge theories, 
particularly at three loops.  Indeed, in any massless gauge theory
the purely-gluonic web diagrams that we need to consider at three 
loops are the same as those arising in $\NeqFour$ super-Yang-Mills theory.
We conclude that at three loops $\Delta$ should have transcendentality 
$\tau = 5$~\cite{Dixon:2009gx}, while for $l > 3$ some relevant 
webs may depend on the matter content of the theory, so that $\Delta$
is only constrained to have a transcendentality at most equal to $2 l - 1$.  

It should be emphasized that some transcendentality could be 
attributed to constant prefactors.  For example, the sum-over-dipoles 
formula~(\ref{Gamma_ansatz}) for $\Gamma_{\dip}$ attains 
transcendentality $\tau = 2 l - 1$ as the sum of $\tau = 2 l - 2$ from 
the (constant) cusp anomalous dimension $\gamma_K$ (associated with
a $1/\e^2$ double pole in the amplitude) and $\tau = 1$ from the single
logarithm. 

Because the functions $\Delta^{\kin}_4$ are defined up to possible 
numerical prefactors, which may carry transcendentality, terms of the 
form~(\ref{kin_4legs}), (\ref{Delta_gen}) or (\ref{Delta_gen_symm}) 
must obey
\begin{equation}
\label{max_trans}
h_{\rm tot} = h_1 + h_2 + h_3 \leq 2 l - 1 \, .
\end{equation}
We note furthermore that constants of transcendentality $\tau = 1$,
{\it i.e.} single factors of $\pi$, do not arise in Feynman diagram
calculations, except for imaginary parts associated with unitarity phases.
We conclude that whenever the maximal transcendentality argument
applies to $\Gamma$, the special case in which our functions
$\Delta_4$ have $\tau = 2 l - 2$ is not allowed.

The sum of the powers of all the logarithms in the 
product must then be no more than $h_{\rm tot} = 5$ at three loops, 
or $h_{\rm tot} = 7$ at four loops, and so on. In the special cases 
considered above, at three loops, the constraint is: $3 h \leq 5$, {\it i.e.} 
$h \leq 1$ in \eqn{Delta_a}, $h \leq 5$ in \eqn{Delta_b}, and $2 h \leq 5$, 
{\it i.e.} $h \leq 2$ in \eqn{Delta_c}. Clearly, at low orders,
transcendentality imposes strict limitations on the admissible
functional forms. We will take advantage of these limitations at three
loops in \sect{sec:3loop}.

We close this subsection by providing some examples of possible
transcendental functions that might enter $\Delta$, beyond the purely
logarithmic examples we have focused on so far.
For functions of multiple dimensionless arguments, the space of
possibilities is not precisely characterized.  Even for kinematical
constants, the allowed structures are somewhat empirically based:
the cusp anomalous dimension, for example, can be expressed through three
loops~\cite{Moch:2004pa} in terms of linear combinations of the 
Riemann zeta values $\zeta(n)$ (having transcendentality $n$),
multiplied by rational numbers; other transcendentals that
might be present --- such as $\ln2$, which does appear in 
heavy-quark mass shifts --- are not.

The cusp anomalous dimension
governs the leading behavior of the twist-two anomalous dimensions 
for infinite Mellin moment $N$.  At finite $N$, these anomalous dimensions
can be expressed~\cite{Moch:2004pa} in terms of the harmonic
sums $S_{\vec{n}_\tau}(N)$~\cite{Vermaseren:1998uu},
where $\vec{n}_\tau$ is a $\tau$-dimensional vector of integers.  
Harmonic sums are the Mellin transforms of harmonic
polylogarithms ${\rm H}_{\vec{m}_\tau}(x)$~\cite{Remiddi:1999ew},
which are generalizations of the ordinary polylogarithms
${\rm Li}_n(x)$.  They are defined recursively by integration,
\begin{equation}
{\rm H}_{\vec{m}_\tau}(x)
\ =\ \int_0^x dx' \ f(a;x') \, {\rm H}_{\vec{m}_{\tau-1}}(x') \,,
\label{harmpolydef}
\end{equation}
where $a=-1$, 0 or 1, and
\begin{equation}
f(-1;x) = {1\over 1+x} \,, \quad 
f(0;x) = {1\over x} \,, \quad 
f(1;x) = {1\over 1-x} \,.
\label{fxdef}
\end{equation}
Note that the transcendentality increases by one unit for each integration.
All three values of $a$ are needed to describe the twist-two anomalous 
dimensions.  However, for the four-point scattering amplitude, which is a 
function of the single dimensionless ratio $r$ defined in \eqn{r_def},
only $a=0,1$ seem to be required~\cite{Smirnov:2003vi}.

Scattering amplitudes depending on two dimensionless ratios can often be
expressed in terms of harmonic polylogarithms as well, but where the
parameter $a$ becomes a function of the second dimensionless
ratio~\cite{Gehrmann:1999as}.  In Ref.~\cite{DelDuca:2009au},
a quantity appearing in a six-point scattering amplitude at two loops
was recently expressed in terms of the closely-related Goncharov
polylogarithms~\cite{Goncharov} in two variables, and at
weight (trancendentality) four.  Other recent works focusing more on 
general mathematical properties include
Refs.~\cite{Bogner:2007mn,Brown:2009rc}.
In general, the space of possible
functions becomes quite large already at weight five, and our examples
below are meant to be illustrative rather than exhaustive.


\subsection{Collinear limits}
\label{CollimitSubsection}

Equipped with the knowledge of how Bose symmetry and other
requirements may be satisfied, let us return to the splitting
amplitude constraint, namely the requirement that the difference
between the two $\Delta$ terms in \eqn{Gamma_Sp_explicit} must
conspire to depend only on the color and kinematic variables of the
two partons that become collinear.

We begin by analyzing the case of an amplitude with precisely four 
colored partons, possibly accompanied by other colorless particles (we 
postpone the generalization to an arbitrary number of partons 
to \sect{sec:n-leg}). The collinear constraint simplifies for $n = 4$ 
because for three partons there are no contributions beyond 
the sum-over-dipoles formula, so that $\Delta_{n - 1} = \Delta_3 = 
0$~\cite{Gardi:2009qi}.\footnote{For $n=4$ we should add a colorless
particle carrying off momentum; otherwise the three-parton kinematics
are ill-defined (for real momenta), and the limit $p_1\cdot p_2 \to 0$
is not really a collinear limit but a forward or
backward scattering limit.}
In~\eqn{Gamma_Sp_explicit} we therefore 
have to consider $\Delta_{4}$ on its own, and require that when, 
say, $p_1$ and $p_2$ become collinear $\Delta_{4}$ does not
involve the kinematic or color variables of other hard particles in
the process.  Because in this limit there remains no non-singular
Lorentz-invariant kinematic variable upon which $\Delta_{4}$ can depend,
it essentially means that $\Delta_{4}$ must become trivial in this limit, 
although it does not imply, of course, that $\Delta_{4}$ vanishes 
away from the limit. In the following we shall see how this can be realized.

To this end let us first carefully examine the limit under consideration.  
We work with strictly massless hard partons, $p_i^2 = 0$ for all $i$. 
In a fixed-angle scattering amplitude we usually consider $2 p_i \cdot 
p_j = Q^2 \beta_i \cdot \beta_j$ where $Q^2$ is taken large,
keeping $\beta_i \cdot \beta_j = {\cal O}(1)$ for any $i$ and $j$. 
Now we relax the fixed-angle limit for the pair of hard partons $p_1$ and 
$p_2$.  Defining $P \equiv p_1 + p_2$ as in \sect{sec:SA}, we
consider the limit $2 p_1 \cdot p_2/Q^2 = P^2/Q^2 \to 0$.
The other Lorentz invariants all remain large; in 
particular for any $j \neq 1,2$ we still have $2 p_1 \cdot p_j = 
Q^2 \beta_1 \cdot \beta_j$ and $2 p_2 \cdot p_j = Q^2 \beta_2
\cdot \beta_j$ where $\beta_1 \cdot \beta_j$ and $\beta_2 \cdot 
\beta_j$ are of ${\cal O}(1)$. In order to control the way in which the 
limit is approached, it is useful to define
\begin{align}
\label{p1_and_p2}
p_1 = z \, P + k \,, \qquad \quad
p_2 = (1 - z) P - k \,,
\end{align}
so that $z$ measures the longitudinal momentum fraction
carried by $p_1$ in $P$, namely
\begin{equation}
\label{z}
z = \frac{p_1^+}{P^+} = \frac{p_1^+}{p_1^+ + p_2^+} \, ,
\end{equation}
where we assume, for simplicity, that the ``$+$'' light-cone
direction\footnote{One can then further specify the frame by choosing 
the ``$-$'' direction along the momentum of one of the other hard 
partons, say $p_3$.} is defined by $p_1$, so that $p_1 = (p_1^+, 0^-, 
0_{\perp})$.  In~\eqn{z} both the numerator and denominator are of 
order $Q$, so $z$ is of ${\cal O}(1)$ and remains fixed in the limit 
$P^2/Q^2 \to 0$. In~\eqn{p1_and_p2} $k$ is a small residual momentum, 
making it possible for $P$ to be off the light-cone while $p_1$ and $p_2$ 
remain strictly light-like. Using the mass-shell conditions $p_1^2 = p_2^2 
= 0$ one easily finds
\begin{equation}
k^2 = - z (1 - z) P^2 \,,
\qquad \quad k \cdot P = \frac12 (1 - 2 z) P^2 \, ,
\end{equation}
so that the components of $k$ are
\begin{equation}
k = \left(0^+, - \frac{P^2}{2P^+}, - \sqrt{z(1 - z)\,P^2} \right) \, .
\end{equation}
Note that in the collinear limit $k^-/Q$ scales as $P^2/Q^2$, while 
$k_\perp/Q$ scales as $\sqrt{P^2/Q^2}$.

We can now examine the behavior of the logarithms of the three cross 
ratios entering $\Delta^{\kin}_4$ in \eqn{kin_4legs}, in the limit $P^2 \to 0$.
Clearly, $L_{1234}$ and $L_{1423}$, which contain the vanishing invariant
$p_1 \cdot p_2$ either in the numerator or in the denominator, will be 
singular in this limit. Similarly, it is easy to see that $L_{1342}$
must vanish, 
because $\rho_{1342} \to 1$. More precisely, the collinear behavior may be
expressed using the parametrization of \eqn{p1_and_p2}, with the result
\begin{align}
\label{p1p2_get_collinear_rho1234}
\begin{split}
L_{1234} &=
\ln \left(\frac{p_1\cdot p_2 \, p_3 \cdot p_4}
{p_1 \cdot p_3 \, p_2 \cdot p_4} \right) \\
&\simeq \,
\underbrace{\ln\left(\frac{P^2 \ p_3 \cdot p_4}
{2 z (1 - z) \, P \cdot p_3 \, P \cdot p_4}\right)}_{
{\cal O} \left(\ln(P^2/Q^2) \right)}
\, - \underbrace{\frac{k \cdot p_3}{z \, P \cdot p_3}}_{{\cal O}
\left(\sqrt{{P^2}/{Q^2}} \right)} 
+ \underbrace{\frac{k \cdot p_4}{(1 - z) \, P \cdot p_4}}_{{\cal O}
\left( \sqrt{{P^2}/{Q^2}} \right)}
\, \to \, \infty \, ,
\end{split}
\\
\label{p1p2_get_collinear_rho1423}
\begin{split}
L_{1423} &=
\ln \left(\frac{p_1 \cdot p_4 \, p_2 \cdot p_3}
{p_1 \cdot p_2 \, p_4 \cdot p_3} \right) \, \\
&\simeq\,
\underbrace{\ln \left(\frac{2 z (1 - z)\ P \cdot p_4 \, P\cdot p_3}
{P^2 \, p_4 \cdot p_3} \right)}_{{\cal O}
\left( \ln(P^2/Q^2) \right)}\,
- \underbrace{\frac{k \cdot p_3}{(1 - z) \, P \cdot p_3}}_{{\cal O}
\left( \sqrt{{P^2}/{Q^2}} \right)} 
+ \underbrace{\frac{k \cdot p_4}{z \, P \cdot p_4}}_{{\cal O}
\left( \sqrt{{P^2}/{Q^2}} \right)}
\, \to \, - \infty \, ,
\end{split}
\\
\label{p1p2_get_collinear_rho1342}
\begin{split}
L_{1342} &=
\ln \left(\frac{p_1 \cdot p_3 \, p_4 \cdot p_2}
{p_1 \cdot p_4 \, p_3 \cdot p_2}\right) \,
= \, \frac{1}{z (1 - z)} \,\left(
\frac{k \cdot p_3}{P \cdot p_3} -
\frac{k \cdot p_4}{P \cdot p_4} \right)
\, = \, {\cal O} \left(\sqrt{{P^2}/{Q^2}} \right) \, \to \, 0 \, ,
\end{split}
\end{align}
where we expanded in the small momentum $k$. As expected, two of the
cross-ratio logarithms diverge logarithmically with $P^2/Q^2$, with
opposite signs, while the third cross-ratio logarithm \emph{vanishes
linearly with} $\sqrt{P^2/Q^2}$. We emphasize that this vanishing is
independent of whether the momenta $p_i$ are incoming or outgoing, except,
of course, that the two collinear partons $p_1$ and $p_2$ must either be
both incoming or both outgoing. Indeed, according to \eqn{rhoijkl_mod},
$\rho_{1342}$ carries no phase when $p_1$ and $p_2$ are collinear:
\begin{equation}
\label{rho1342_phase}
 \rho_{1342} 
=\left|\frac{p_1 \cdot p_3 \ p_4 \cdot p_2}
        {p_1 \cdot p_4 \ p_3 \cdot p_2} \right| 
{\rm e}^{-{\rm i}\pi(\lambda_{13} + \lambda_{42} 
                  -  \lambda_{14} - \lambda_{32})} 
\to 1\, ,
\end{equation}
since $\lambda_{13}=\lambda_{32}$ and $\lambda_{42}=\lambda_{14}$.

Let us now examine a generic term with a kinematic dependence 
of the form~(\ref{kin_4legs}) in this limit. Substituting 
eqs.~(\ref{p1p2_get_collinear_rho1234}) through 
(\ref{p1p2_get_collinear_rho1342}) into \eqn{kin_4legs} we see 
that, if $h_3$ (the power of $L_{1342}$) is greater than or equal 
to $1$, then the result for $\Delta^{\kin}_4$ in the collinear limit 
is zero. This vanishing is not affected by the powers of the other 
logarithms, because they diverge only logarithmically as $P^2/Q^2 
\to 0$, while $L_{1342}$ vanishes as a power law in the same limit.  
In contrast, if $h_3 = 0$, and $h_1$ or $h_2$ is greater than
zero, then the kinematic function $\Delta^{\kin}_4$ in \eqn{kin_4legs} 
diverges when $p_1$ and $p_2$ become collinear, due to the behavior 
of $L_{1234}$ and $L_{1423}$ in 
\eqns{p1p2_get_collinear_rho1234}{p1p2_get_collinear_rho1423}.
The first term in each of these equations introduces explicit 
dependence on the non-collinear parton momenta $p_3$ and $p_4$
into $\Delta_n$ in \eqn{Gamma_Sp_explicit}, which would violate
collinear universality. We conclude that consistency with the limit 
where $p_1$ and $p_2$ become collinear requires $h_3 \geq 1$.

Obviously we can consider, in a similar way, the limits where other
pairs of partons become collinear, leading to the conclusion that all
three logarithms, $L_{1234}$, $L_{1423}$ and $L_{1324}$ must appear 
raised to the first or higher power. Collinear limits thus constrain
the powers of the logarithms by imposing 
\begin{equation}
h_i \geq 1 \, , \qquad  \forall \, i \, . 
\label{splitting_ampl_constraint}
\end{equation}
This result puts a lower bound on the 
transcendentality of~$\Delta^{\rm kin}_4$, namely
\begin{equation}
\label{min_trans}
h_{\rm tot} = h_1 + h_2 + h_3 \geq 3 \, .
\end{equation}


\subsection{Three-loop analysis\label{sec:3loop}}

We have seen that corrections to the sum-over-dipoles formula involving 
four-parton correlations are severely constrained. We can now examine
specific structures that may arise at a given loop order $l$, beginning 
with the first nontrivial possibility, $l = 3$. Because we consider webs that 
are attached to four hard partons, at three loops they can only attach 
once to each eikonal line, as in \fig{4Elabeled}, giving the color factor 
in \eqn{color}, where $h^{abcd}$ must be constructed out of the structure 
constants $f^{ade}$. The only possibility is terms of the form $f^{ade} 
f^{bce}$ --- the same form we obtained in the previous section starting 
from the symmetry properties of the kinematic factors depending on 
$L_{ijkl}$. In contrast, the symmetric tensor $d^{ade}$ cannot arise in 
three-loop webs.

Taking into account the splitting amplitude 
constraint~(\ref{splitting_ampl_constraint}) on the one hand,
and the maximal transcendentality constraint~(\ref{max_trans})
on the other, there are just a few possibilities for the various powers 
$h_i$. These are summarized in Table~\ref{table:_hi}.

The lowest allowed transcendentality for $\Delta_4^{\rm kin}$
is $\tau = 3$, corresponding to $h_1 = h_2 = h_3 = 1$. This brings 
us to \eqn{Delta_a}, in which we would have to construct
a completely antisymmetric tensor $h^{abcd}$ out of the structure 
constants $f^{ade}$.  Such a tensor, however, does not exist. 
Indeed, starting with the general expression~(\ref{Delta_gen}), which 
is written in terms of the structure constants, and substituting $h_1 = 
h_2 = h_3 = 1$, we immediately see that the color structure factorizes, 
and vanishes by the Jacobi identity~(\ref{Jacobi}). The possibility $h_1 = 
h_2 = h_3 = 1$ is thus excluded by Bose symmetry.

Next, we may consider transcendentality $\tau = 4$. Ultimately, we 
exclude functions with this degree of transcendentality at three loops,
because we are dealing with purely gluonic webs, which are the same
as in ${\cal N} = 4$ super Yang-Mills theory. We expect then that the 
anomalous dimension matrix will have a uniform degree of transcendentality
$\tau = 5$, and there are no constants with $\tau = 1$ that might
multiply functions with $h_{\rm tot} =  4$ to achieve the desired result, as 
discussed in \sect{sec:maxtran}.  However, it is instructive to note that
symmetry alone does not rule out this possibility. Indeed, having 
excluded \eqn{Delta_gen_symm}, involving the symmetric tensor 
$d^{ade}$, we may consider \eqn{Delta_gen} with $h_1 + h_2 + 
h_3 = 4$. Bose symmetry and the splitting amplitude constraint in 
\eqn{splitting_ampl_constraint} leave just two potential structures, 
one with $h_1 = 2$ and $h_2 = h_3 = 1$, and a second one with
$h_1 = h_2 = 1$ and $h_3 = 2$ ($h_1 = h_3 = 1$ and $h_2 = 2$ 
yields the latter structure again). The former vanishes identically, while
the latter could provide a viable candidate, 
\begin{align}
\label{Delta_112}
\begin{split}
&\Delta_4^{(112)} (\rho_{ijkl}) \, = \, 
\, {\bf T}_1^{a} {\bf T}_2^{b} {\bf T}_3^{c} {\bf T}_4^{d} \\
&\hskip1cm
\times \Big[f^{ade}f^{cbe}
L_{1234}\,\Big( L_{1423}\, L_{1342}^{2}
- \, L_{1342} \, L_{1423}^{2}\Big)
\,  \\
&\hskip1.5cm
+ f^{cae}f^{dbe}
L_{1423}\,\Big( L_{1342}\, L_{1234}^2
- \, L_{1234} \, L_{1342}^{2}\Big) \\
&\hskip1.5cm
+ f^{bae}f^{cde}
L_{1342}\, \Big(L_{1234}\, L_{1423}^{2}
- \, L_{1423}\, L_{1234}^{2}\Big) 
\Big] \,.
\end{split}
\end{align}
We rule out \eqn{Delta_112} based only on its degree of transcendentality.
\begin{table}[htb]
\begin{center}
\begin{tabular}{|c|c|c|l|l l l|}
\hline
$h_1$&$h_2$&$h_3$& $h_{\rm tot}$ & comment&&\\
\hline
 1   &   1 &   1 &  3            & vanishes identically by Jacobi
 identity~(\ref{Jacobi})&&\\
 2   &   1 &   1 &  4            & kinematic factor vanishes identically&&\\
 1   &   1 &   2 &  4            & allowed by symmetry, excluded by
transcendentality&& \\
 1   &   2 &   2 &  5            & viable possibility, \eqn{Delta_case122}&
\multirow{4}{*}{\hspace*{-130pt} {\fontsize{54}{15}\selectfont $\}$}} &
\multirow{4}{*}{\hspace*{-110pt} all coincide using \eqn{Jacobi} }\\
 3   &   1 &   1 &  5            & viable possibility, \eqn{Delta_case311}&&\\
 2   &   1 &   2 &  5            & viable possibility, \eqn{Delta_case212}&&\\
 1   &   1 &   3 &  5            & viable possibility, \eqn{Delta_case113}&&\\
\hline
\end{tabular}
\end{center}
\caption{Different possible assignments of the powers $h_i$ in
\eqn{Delta_gen} at three loops. We only consider $h_i\geq1$ because
of the splitting amplitude
constraint~(\ref{splitting_ampl_constraint}) and
$h_{\rm tot}\leq 5$ because of the bound on transcendentality,
\eqn{max_trans}.  We also omit the combinations that can be obtained
by interchanging the values of $h_2$ and $h_3$; 
this interchange yields the same function,
up to a possible overall minus sign.\label{table:_hi}}
\end{table}

We consider next the highest attainable transcendentality at three loops, 
$\tau = 5$.  \Eqn{Delta_gen} yields four different structures,
summarized in Table~\ref{table:_hi}. The first structure we consider
has $h_1 = 1$ and $h_2 = h_3 = 2$. It is given by
\begin{align}
\label{Delta_case122}
\begin{split}
&\Delta_4^{(122)} (\rho_{ijkl}) \, = \,
\, {\bf T}_1^{a} {\bf T}_2^{b} {\bf T}_3^{c} {\bf T}_4^{d}
\Big[ f^{ade} f^{cbe}
L_{1234} \, (L_{1423}\,L_{1342})^{2}\\&
+
f^{cae} f^{dbe}
L_{1423}\, (L_{1234}\, L_{1342})^{2}
+
f^{bae} f^{cde}
L_{1342}\, (L_{1423}\, L_{1234})^{2}
\Big] \, .
\end{split}
\end{align}
The second structure has $h_1 = 3$ and $h_2 = h_3 = 1$, yielding
\begin{align}
\label{Delta_case311}
\begin{split}
&\Delta_4^{(311)}(\rho_{ijkl}) \, = \,
\, {\bf T}_1^{a} {\bf T}_2^{b} {\bf T}_3^{c} {\bf T}_4^{d}
\Big[f^{ade} f^{cbe}
(L_{1234})^{3}\, L_{1423}\, L_{1342}\\&
+
f^{cae} f^{dbe}
(L_{1423})^{3}\,L_{1234}\, L_{1342}
+
f^{bae} f^{cde}
(L_{1342})^{3}\, L_{1423}\, L_{1234}
\Big] \, .
\end{split}
\end{align}

We now observe that the two functions~(\ref{Delta_case122}) 
and (\ref{Delta_case311}) are, in fact, one and the same.  To show
this, we form their difference, and use relation~(\ref{three_rho_relation}) 
to substitute $L_{1234} = - L_{1423} \, - \, L_{1342}$.  We obtain
\begin{align}
\label{122_and_311_combination}
\begin{split}
\Delta_4^{(122)} - \Delta_4^{(311)} \, =& 
\ {\bf T}_1^{a} {\bf T}_2^{b} {\bf T}_3^{c} {\bf T}_4^{d}
 \,\, L_{1234} \, L_{1423} \, L_{1342}
\\& \hskip0.1cm \times \bigg[f^{ade}f^{cbe}
\Big( L_{1423}\,L_{1342} \,-\, L_{1234}^{2}\Big)
\, + \,
f^{cae}f^{dbe}
\Big( L_{1234}\, L_{1342} \,-\, L_{1423}^{2}\Big)
\\& \hskip0.5cm +
f^{bae}f^{cde}
\Big(L_{1423}\, L_{1234} \,-\, L_{1342}^{2} \Big)
\bigg]
\\ =&
\, - \,  {\bf T}_1^{a} {\bf T}_2^{b} {\bf T}_3^{c} {\bf T}_4^{d}
 \,\, L_{1234} \, L_{1423} \, L_{1342}\ 
\left[f^{ade}f^{cbe}+ f^{cae}f^{dbe}+f^{bae}f^{cde} \right] 
\\ & \hskip0.1cm \times 
\left( L_{1342}^2 \, + \, L_{1342} \, L_{1423} \, 
 + \, L_{1423}^{2} \right) \, = \, 0 \, ,
\end{split}
\end{align}
vanishing by the Jacobi identity~(\ref{Jacobi}).

The last two structures in Table~\ref{table:_hi} are given by 
\begin{align}
\label{Delta_case212}
\begin{split}
\Delta_4^{(212)} (\rho_{ijkl}) \, = \,
\, {\bf T}_1^{a} {\bf T}_2^{b} {\bf T}_3^{c} {\bf T}_4^{d}
\Big[ &f^{ade} f^{cbe}
L_{1234}^2 \, \Big(L_{1423}\,L_{1342}^{2}\,+
\,L_{1423}^{2}\,L_{1342}\Big) \\& +
f^{cae} f^{dbe}
L_{1423}^2\, \Big(L_{1234}\, L_{1342}^{2}\, +\, 
L_{1234}^{2}\, L_{1342} \Big) \\& +
f^{bae} f^{cde}
L_{1342}^{2}\, \Big(L_{1423}\, L_{1234}^{2} \,+\,
L_{1423}^{2} \, L_{1234}\Big)
\Big] \, ,
\end{split}
\end{align}
and
\begin{align}
\label{Delta_case113}
\begin{split}
\Delta_4^{(113)}(\rho_{ijkl}) \, = \,
\, {\bf T}_1^{a} {\bf T}_2^{b} {\bf T}_3^{c} {\bf T}_4^{d}
\Big[&f^{ade} f^{cbe}
L_{1234}\, \Big(L_{1423}\, L_{1342}^{3}\,+\,
L_{1423}^{3}\, L_{1342}\Big) \\& +
f^{cae} f^{dbe}
L_{1423}\,\Big(L_{1234}\, L_{1342}^{3}\,+\,
L_{1234}^{3}\, L_{1342}\Big) \\& + 
f^{bae} f^{cde}
L_{1342}\, \Big(L_{1423}\, L_{1234}^{3}\,+\,
L_{1423}^{3}\, L_{1234}\Big) \Big] \, .
\end{split}
\end{align}
One easily verifies that they are both proportional to $\Delta_4^{(122)} = 
\Delta_4^{(311)}$. Consider first \eqn{Delta_case212}.  In each line we 
can factor out the logarithms and use \eqn{three_rho_relation}
to obtain a monomial. For example, the first line may be written as:
\begin{align}
\begin{split}
L_{1234}^2\,\Big( L_{1423}\,L_{1342}^{2}\,+\,L_{1423}^{2}\,L_{1342}\Big) 
&=\,L_{1234}^2\,L_{1423}\,L_{1342}\, (L_{1423}+L_{1342}) \\
&=\, - L_{1234}^3 \,L_{1423}\,L_{1342}\, ,
\end{split}
\end{align}
where we recognise that this function coincides with \eqn{Delta_case311} above.
Consider next \eqn{Delta_case113}, where, for example, the first line yields
\begin{align}
\begin{split}
L_{1234}\,\Big(L_{1423}\, L_{1342}^{3}\,+\,L_{1423}^{3}\, L_{1342}\Big) &=
L_{1234}\,L_{1423}\, L_{1342} \Big(L_{1342}^{2}\,+\,L_{1423}^{2}\Big)
\\&=L_{1234}\,L_{1423}\, L_{1342} 
\Big(\left(L_{1342}\,+\,L_{1423}\right)^{2}-2L_{1342}\,L_{1423}\Big)\,
\\&=L_{1234}\,L_{1423}\, L_{1342} \Big(L_{1234}^{2}-2L_{1342}\,L_{1423}\Big)\,,
\end{split}
\end{align}
which is a linear combination of \eqns{Delta_case122}{Delta_case311},
rather than a new structure.

We conclude that there is precisely one function, 
$\Delta_4^{(122)} = \Delta_4^{(311)}$, that can be 
constructed out of arbitrary powers of logarithms and is consistent 
with all available constraints at three loops.
We emphasize that this function is built with color and kinematic
factors that one expects to find in the actual web diagram 
computations, and it is quite possible that it indeed appears.
Because this structure saturates the transcendentality bound, its 
coefficient is necessarily a rational number.  

Note that color conservation has not been imposed here, but
it is implicitly assumed that for a four-parton amplitude
\begin{equation}
{\bf T}_1^{a} \,+\, {\bf T}_2^{a}
\,+\, {\bf T}_3^{a} \,+\, {\bf T}_4^{a} = 0 \, .
\end{equation}
Importantly, upon using this relation, the structure~(\ref{Delta_case122}) 
(or, equivalently, (\ref{Delta_case311})) remains non-trivial.


\subsection{Three-loop functions involving polylogarithms\label{sec:poly}}

Additional functions can be constructed upon removing the requirement
that the kinematic dependence be of the form~(\ref{kin_4legs}),
where only powers of logarithms are allowed.  Three key 
features of the function $\ln\rho$ were essential in the examples above:
it vanishes like a power at $\rho = 1$, it has a definite symmetry under 
$\rho \to 1/\rho$, and it only diverges logarithmically as
$\rho \to 0$ and $\rho \to \infty$.  These properties can be mimicked by
a larger class of functions.  In particular, allowing dilogarithms one
can easily
construct a function of transcendentality $\tau = 4$, which is consistent 
with Bose symmetry and collinear constraints. It is given by
\begin{eqnarray}
\label{Delta_case211_dilog}
\Delta_4^{(211,\, {\rm Li_2})} (\rho_{ijkl}) &  = &
\,{\bf T}_1^{a} {\bf T}_2^{b} {\bf T}_3^{c} {\bf T}_4^{d}\\
&& \bigg[ f^{ade}f^{cbe}
\Big({\rm Li}_2(1-\rho_{1234}) - {\rm Li}_2(1-1/\rho_{1234})\Big)
\, \ln\rho_{1423}\ \ln\rho_{1342}
\nonumber \\ && + f^{cae}f^{dbe}
\Big({\rm Li}_2(1-\rho_{1423}) - {\rm Li}_2(1-1/\rho_{1423})\Big)
\ln\rho_{1234}\ \ln\rho_{1342}
\nonumber \\ && + f^{bae}f^{cde}
\Big({\rm Li}_2(1-\rho_{1342}) - {\rm Li}_2(1-1/\rho_{1342})\Big)
  \,\ln\rho_{1423}\ \ln\rho_{1234}
\bigg] \,. \nonumber 
\end{eqnarray}
The key point here is that the function ${\rm Li}_2(1 - \rho_{1234}) - 
{\rm Li}_2(1 - 1/\rho_{1234})$ is odd under $\rho_{1234} \to 
1/\rho_{1234}$, which allows it to be paired with the antisymmetric
structure constants $f^{ade}$.  It is also easy to verify that the
collinear constraints are satisfied.

We note that it is also possible to construct a potentially relevant function 
containing logarithms with a more complicated kinematic dependence.
Indeed, the structure
\begin{eqnarray}
\label{Delta_case211_mod}
\Delta_4^{(211, \, {\rm mod})} (\rho_{ijkl}) & = &
\, {\bf T}_1^{a} {\bf T}_2^{b} {\bf T}_3^{c} {\bf T}_4^{d}
\\ && \Bigg[f^{ade}f^{cbe}
\ln \rho_{1234}\ \ln\left(\frac{\rho_{1234}}
{(1 - \rho_{1234})^2}\right) \,\ln\rho_{1423}\ \ln\rho_{1342}
\nonumber \\ && +
f^{cae}f^{dbe}
\ln \rho_{1423}\ \ln\left(\frac{\rho_{1423}}
{(1 - \rho_{1423})^2}\right)\, \ln\rho_{1234}\ \ln\rho_{1342}
\nonumber \\ && +
f^{bae}f^{cde}
\ln \rho_{1342}\,\,\ln\left(\frac{\rho_{1342}}
{(1 - \rho_{1342})^2}\right) \,\ln\rho_{1423}\ \ln\rho_{1234}
\Bigg] \,  \nonumber  
\end{eqnarray}
fulfills the symmetry requirements discussed above, because
$\ln\left({\rho_{1234}}/{(1 - \rho_{1234})^2}\right)$ is even under
$\rho_{1234} \to 1/\rho_{1234}$. Thanks to the extra power of the
cross-ratio logarithm, eq.~(\ref{Delta_case211_mod}) also vanishes in all
collinear limits, as required.  Logarithms with argument $1 - \rho_{ijkl}$
cannot be directly rejected on the basis of the fact that they induce
unphysical singularities, because $\rho_{ijkl} \to 1$ corresponds to a
physical collinear limit\footnote{Note that the analogous structure
containing $\ln \left({\rho_{1234}}/{(1 + \rho_{1234})^2}\right)$ can be
excluded because the limit $\rho_{ijkl} \to -1$ should not be
singular. Indeed, by construction the variables $\rho_{ijkl}$ always
contain an even number of negative momentum invariants, so their real part
is always positive (although unitarity phases may add up and bring their
logarithm to the second Riemann sheet).}.  We conclude that
\eqns{Delta_case211_dilog}{Delta_case211_mod} would be viable based on
symmetry and collinear requirements alone.  However, we can exclude them
on the basis of transcendentality: as discussed in \sect{sec:maxtran}, a
function with $h_{\rm tot} = 4$ cannot arise at three loops, because it
cannot be upgraded to maximal transcendentality $\tau=5$ by constant
prefactors.

At transcendentality $\tau = 5$, there are at least two further viable
structures that involve polylogarithms, in which second and
third powers of logarithms are replaced, respectively, by appropriate
combinations ${\rm Li}_2$ and ${\rm Li}_3$. The first structure can be obtained
starting from \eqn{Delta_case122}, and using the same combination of 
dilogarithms that was employed in \eqn{Delta_case211_dilog}. One finds
\begin{eqnarray}
\label{Delta_case122_mod}
&& \hspace{5mm} \Delta_4^{(122, \, {\rm Li_2})} (\rho_{ijkl}) \, = \,
\,{\bf T}_1^{a} {\bf T}_2^{b} {\bf T}_3^{c} {\bf T}_4^{d}
\\ && \hspace{-3mm} \times \bigg[f^{ade}f^{cbe}
\ln\rho_{1234}\, 
\Big({\rm Li}_2(1 - \rho_{1342}) - {\rm Li}_2 (1 - 1/\rho_{1342}) \Big)
\Big({\rm Li}_2(1 - \rho_{1423}) - {\rm Li}_2 (1 - 1/\rho_{1423}) \Big) 
\nonumber \\ && \hspace{-3mm} +  f^{cae}f^{dbe}
\ln\rho_{1423}\, 
\Big({\rm Li}_2(1-\rho_{1234}) - {\rm Li}_2(1-1/\rho_{1234})\Big)
\Big({\rm Li}_2(1-\rho_{1342}) - {\rm Li}_2(1-1/\rho_{1342})\Big) 
\nonumber \\ && \hspace{-3mm} +  f^{bae}f^{cde}
\ln\rho_{1342}\, 
\Big({\rm Li}_2(1-\rho_{1234}) - {\rm Li}_2(1-1/\rho_{1234})\Big)
\Big({\rm Li}_2(1-\rho_{1423}) - {\rm Li}_2(1-1/\rho_{1423})\Big) 
\bigg] . \nonumber
\end{eqnarray}
Here it was essential to replace both $\ln^2$ terms in order to keep
the symmetry properties in place.  Starting instead from
\eqn{Delta_case311}, there is one possible polylogarithmic
replacement, which, however, requires introducing trilogarithms,
because using ${\rm Li}_2$ times a logarithm would turn the odd
function into an even one, which is excluded. One may write instead
\begin{eqnarray}
\label{Delta_case311_mod}
\Delta_4^{(311, \, {\rm Li_3})} (\rho_{ijkl}) & = &
\,{\bf T}_1^{a} {\bf T}_2^{b} {\bf T}_3^{c} {\bf T}_4^{d}\\
&& \times \bigg[f^{ade}f^{cbe}
\Big({\rm Li}_3(1-\rho_{1234}) - {\rm Li}_3(1-1/\rho_{1234})\Big)
 \, L_{1423}\, L_{1342}
\nonumber \\
&& \hskip0.4cm +f^{cae}f^{dbe}
\Big({\rm Li}_3(1-\rho_{1423}) - {\rm Li}_3(1-1/\rho_{1423})\Big)
 \,L_{1234}\, L_{1342}
\nonumber \\
&& \hskip0.4cm +f^{bae}f^{cde}
\Big({\rm Li}_3(1-\rho_{1342}) - {\rm Li}_3(1-1/\rho_{1342})\Big)
\, L_{1423}\, L_{1234}
\bigg] \, . \nonumber 
\end{eqnarray}
Neither \eqn{Delta_case122_mod} nor \eqn{Delta_case311_mod} can be 
excluded at present, as they satisfy all available constraints.
We can, however, exclude similar constructions with higher-order 
polylogarithms.  For example, ${\rm Li}_4$ has transcendentality 
$\tau = 4$, so it could be accompanied by at most one logarithm; this 
product would not satisfy all collinear constraints.  
We do not claim to be exhaustive in our investigation of 
polylogarithmic functions; additional possibilities may arise upon allowing 
arguments of the polylogarithms that have a different functional
dependence on the cross ratios.


\subsection{Four-loop analysis}

Let us briefly turn our attention to contributions that may arise beyond 
three loops. At the four-loop level several new possibilities open up. First, 
there are potential quartic Casimirs in $\gamma_K$. Corresponding 
corrections to the soft anomalous dimension would satisfy
inhomogeneous differential equations, eq. (5.5) of Ref.~\cite{Gardi:2009qi}.
Beyond that, new types of corrections may appear even if $\gamma_K$ 
admits Casimir scaling. First, considering the logarithmic expressions of
\eqn{kin_4legs}, purely gluonic webs might give rise to functions
of transcendentality up to $h_{\rm tot} = h_1 + h_2 + h_3 = 7$. 
At this level, there are four potential functions: (a) $h_1 = 5$ and 
$h_2 = h_3 = 1$; (b) $h_1 = 4$, $h_2 = 2$, $h_3 = 1$; (c) $h_1 = 1$ 
and $h_2 = h_3 = 3$; (d) $h_1 = 3$ and $h_2 = h_3 = 2$. 
Of course, as in the three-loop case, also polylogarithmic structures may 
appear, and functions with $h_{\rm tot} \leq 5$ might be present (of the
type already discussed at three loops), multiplied by transcendental
constants with $\tau\geq2$.

It is interesting to focus in particular on color structures that are related
to quartic Casimir operators, which can appear at four loops not only in 
$\gamma_K$ but also in four-parton correlations. Indeed, a structure 
allowed by Bose symmetry and collinear constraints is given by
\eqn{Delta_a}, where the group theory factor $h^{abcd}$ is generated
by a pure-gluon box diagram attached to four different hard partons, giving 
rise to a trace of four adjoint matrices. It is of the form
\begin{equation}
\label{four_loop_example_adjoint}
\Delta_4^{C_{4, A}} (\rho_{ijkl}) \, = \, 
{\rm Tr}\left[F^{a}F^{b}F^{c}F^{d}\right]
\,{\bf T}_1^{a} {\bf T}_2^{b} {\bf T}_3^{c} {\bf T}_4^{d}
\,\,\Big[ \ln\rho_{1234} \, \ln\rho_{1423} \, \ln\rho_{1342} 
\Big]^{h} \, ,
\end{equation}
where $F^a$ are the ${\rm SU}(N_c)$ generators in the adjoint 
representation, $(F^a)_{bc} = - {\rm i} f^a_{\, \, \, bc}$. This 
expression may be relevant {\it a priori} for both odd and even $h$, 
projecting respectively on the totally antisymmetric or symmetric 
parts of ${\rm Tr}\left[F^{a}F^{b}F^{c}F^{d}\right]$. As noted 
above, however, a totally antisymmetric tensor cannot be constructed 
with four adjoint indices, so we are left with the completely symmetric 
possibility, which indeed corresponds to the quartic Casimir operator.
The transcendentality constraint comes into play here: the only even
integer $h$ that can give transcendentality $\tau \leq 7$ is $h = 2$.
For $\NeqFour$ super-Yang-Mills theory, \eqn{four_loop_example_adjoint}
with $h=2$ can be excluded at four loops, because there is no 
numerical constant with $\tau = 1$ that could bring the
transcendentality of \eqn{four_loop_example_adjoint} from 6 up to 7.

On the other hand, in theories with a lower number of supersymmetries,
and at four loops, there are potentially both pure-glue and matter-loop 
contributions of lower transcendentality, because only the specific loop 
content of $\NeqFour$ super-Yang-Mills theory is expected to be purely 
of maximal transcendentality. Thus \eqn{four_loop_example_adjoint} may 
be allowed for $h=2$ for generic adjoint-loop contributions (for example 
a gluon box in QCD), and analogously
\begin{equation}
\label{four_loop_example_fundm}
\Delta_4^{C_{4, F}} (\rho_{ijkl}) \, = \, 
{\rm Tr}\left[t^{a}t^{b}t^{c}t^{d}\right]
\,{\bf T}_1^{a} {\bf T}_2^{b} {\bf T}_3^{c} {\bf T}_4^{d}
\,\,\Big[ \ln\rho_{1234} \, \ln\rho_{1423} \, \ln\rho_{1342} 
\Big]^{h} \,,
\end{equation}
for loops of matter in the fundamental representation (with
generators $t^a$), {\it e.g.} from quark box diagrams.
As before, the other power allowed by
transcendentality, $h=1$, is excluded by symmetry,
because there is no projection of 
${\rm Tr}\left[t^{a}t^{b}t^{c}t^{d}\right]$ 
that is totally antisymmetric under permutations.

While \eqn{four_loop_example_adjoint} is excluded by transcendentality
for $\NeqFour$ super-Yang-Mills theory, another construction involving
the quartic Casimir is allowed: \eqn{Delta_gen_symm_alt}
for $h_1 = 2$, $h_2 = h_3 = 1$ can be used, after multiplying it by the
transcendentality $\tau = 3$ constant $\zeta(3)$.  Finally, as already 
mentioned, there are a number of other purely logarithmic structures 
with partial symmetry in each term, as represented
by~\eqns{Delta_gen}{Delta_gen_symm}, that may appear at four loops.


\section{Generalization to $n$-leg amplitudes\label{sec:n-leg}}

The above analysis focused on the case of four partons, because this is
the first case where cross ratios can be formed, and thus $\Delta$ may 
appear.  However, rescaling-invariant ratios involving more than four
momenta can always be split into products of cross ratios involving
just four momenta.  Therefore it is straightforward to generalize 
the results we obtained to any $n$-parton process at three loops.
Indeed, contributions to
$\Delta_n$ are simply constructed as a sum over all possible sets of
four partons,
\begin{equation}
\label{Delta_n_legs}
\Delta_n = \sum_{i,j,k,l} \Delta_4(\rho_{ijkl}) \, , 
\end{equation}
just as the sum-over-dipoles formula~(\ref{barS_ansatz}) is written as
a sum over all possible pairs of legs. The indices in the sum in
\eqn{Delta_n_legs} are of course all unequal.  Assuming a purely
logarithmic structure, at three loops the function $\Delta_4$ in
\eqn{Delta_n_legs} is given by $\Delta_4^{(122)}$ in \eqn{Delta_case122}
(or, equivalently, $\Delta_4^{(311)}$ in \eqn{Delta_case311}).
Of course the overall prefactor to $\Delta_4$ could still be zero;
its value remains to be determined by an explicit computation.
The total number of terms in the
sum increases rapidly with the number of legs: for $n$ partons, there 
are $({n \atop 4})$ different terms.

Now we wish to show that this generalization is a consistent one.
To do so, we shall verify that the difference between $\Delta_n$ and
$\Delta_{n - 1}$ in \eqn{Gamma_Sp_explicit},
for the splitting amplitude anomalous dimension,
does not introduce any dependence on the kinematics or color of 
any partons other than the collinear pair.
The verification is non-trivial for $n\geq5$ because
$\Delta_{n - 1}$ is no longer zero.

Consider the general $n$-leg amplitude, in which the two legs $p_1$
and $p_2$ become collinear. The terms entering
\eqn{Gamma_Sp_explicit} include:
\begin{itemize}
\item{} A sum over $({n - 2 \atop 4})$ different terms in $\Delta_n$ that
do not involve any of the legs that become collinear. They depend on
the cross ratios $\rho_{ijkl}$  where none of the indices is $1$ or $2$.
However, exactly the same terms appear in $\Delta_{n - 1}$, so 
they cancel in \eqn{Gamma_Sp_explicit}.
\item{} A sum over $({n - 2 \atop 2})$ different terms in $\Delta_n$
depending on the variables $\rho_{12ij}$ (and permutations), where
$i, \, j \neq 1, \, 2$. These variables involve the two legs that become
collinear.  According to \eqn{Delta_n_legs}, each of these
terms is $\Delta_4$, namely it is given by a sum of terms that admit the
constraint~(\ref{splitting_ampl_constraint}).  Therefore each of them is
guaranteed to vanish in the collinear limit, and we can discard them
from \eqn{Gamma_Sp_explicit}.  The same argument applies to any
$\Delta_4$ that is consistent at the four-parton level, such as the 
polylogarithmic constructions~(\ref{Delta_case122_mod}) and
(\ref{Delta_case311_mod}).
\item{} Finally, $\Delta_n$ brings a sum over 
$2 \times ({n - 2 \atop 3})$ terms involving just one leg among
the two that become collinear.
These terms depend on $\rho_{1jkl}$ or $\rho_{2jkl}$,
where $j, \, k, \, l \neq 1, \, 2$. In contrast, $\Delta_{n - 1}$ brings just
one set of such terms, because the $(12)$ leg, $P = p_1 + p_2$, is now
counted once.  Recalling, however, that this leg carries the color
charge
\begin{equation}
\label{color_consrv_12}
{\bf T}^{a}\,=\,{\bf T}_1^{a} \,+\, {\bf T}_2^{a} \, ,
\end{equation} 
it becomes clear that any term of this sort having a color factor of
the form~(\ref{color}) would cancel out in the difference. Indeed
\begin{align}
\label{collmixed}
\begin{split}
&\Delta_n \left(\rho_{1 j k l}, \rho_{2 j k l} \right) 
-  \Delta_{n - 1} \left(\rho_{(12) j k l} \right) \\
& \hspace{-2mm} = \,
\sum_{j,k,l} \,  h^{abcd} \, {\bf T}_j^{b} {\bf T}_k^{c} 
{\bf T}_l^{d} \,
\Bigg[ \underbrace{\,{\bf T}_1^{a} \Delta^{\kin}(\rho_{1jkl})
+ \,{\bf T}_2^{a} 
\, \Delta^{\kin}(\rho_{2jkl})}_{\text{from}\, \Delta_n} 
- \underbrace{ \,{\bf T}^{a} 
\, \Delta^{\kin}(\rho_{(12) j k l})}_{\text{from}\, \Delta_{n - 1}} \Bigg]
\\
& \hspace{-2mm} = \, 0 \, .
\end{split}
\end{align} 
To show that this combination vanishes we used \eqn{color_consrv_12}
and the fact that the kinematic factor $\Delta^{\kin}$ in all
three terms is identical because of rescaling invariance, that is,
it depends only on the directions of the partons, which coincide
in the collinear limit.
\end{itemize}
We conclude that \eqn{Delta_n_legs} is consistent with the limit as
any two of the $n$ legs become collinear.

A similar analysis also suggests that \eqn{Delta_n_legs} is
consistent with the triple collinear limit in which $p_1$, $p_2$ and $p_3$
all become parallel.  We briefly sketch the analysis.
We assume that there is a universal factorization
in this limit, in which the analog of ${\bf Sp}$ again only depends on the
triple-collinear variables: $P^2\equiv (p_1+p_2+p_3)^2$, which vanishes
in the limit; $2p_1\cdot p_2/P^2$ and $2p_2\cdot p_3/P^2$;
and the two independent longitudinal momentum fractions for the $p_i$, namely
$z_1$ and $z_2$ (and $z_3=1-z_1-z_2$) --- see {\it e.g.} 
Ref.~\cite{Catani:2003vu} for a discussion at one loop.
In the triple-collinear limit there are the following types of
contributions:
\begin{itemize}
\item $({n-3 \atop 4})$ terms in $\Delta_n$
that do not involve any of the collinear legs.  They cancel
in the analog of~\eqn{Gamma_Sp_explicit} between $\Delta_n$ and
$\Delta_{n-2}$, exactly as in the double-collinear case.
\item $3 \times ({n - 3 \atop 3})$ terms containing cross ratios
of the form $\rho_{1jkl}$, or similar terms with 1 replaced by 2 or 3.
These contributions cancel exactly as in \eqn{collmixed}, except 
that there are three terms and the color conservation equation is
${\bf T}^{a}={\bf T}_1^{a}+{\bf T}_2^{a}+{\bf T}_3^{a}$.
\item $3 \times ({n - 3 \atop 2})$ terms containing cross ratios
of the form $\rho_{12kl}$, or similar terms with $\{1,2,3\}$ permuted.
These terms cancel for the same reason as the $\rho_{12ij}$ terms 
in the double-collinear analysis, namely one of the logarithms is 
guaranteed to vanish.
\item $(n-3)$ terms containing cross ratios of the form $\rho_{123l}$.
This case is non-trivial, because no logarithm vanishes (no cross ratio
goes to 1).  However, it is easy to verify that in the limit, each of the 
cross ratios that appears depends only on the triple-collinear
kinematic variables, and in a way that is independent of $p_l$.
Therefore the color identity
$\sum_{l\neq1,2,3} {\bf T}_l^{a} = - {\bf T}^{a}$
can be used to express the color, as well as the kinematic dependence,
of the limit of $\Delta_n$ solely in terms of the collinear variables,
as required by universality.
\end{itemize}
Thus all four contributions are consistent with a universal
triple-collinear limit.  However, because the last type of contribution
is non-vanishing in the limit, in contrast to the
double-collinear case, the existence of a non-trivial $\Delta_n$
would imply a new type of contribution to the $1/\epsilon$ pole
in the triple-collinear splitting function, beyond that implied by the
sum-over-dipoles formula.

We conclude that \eqn{Delta_n_legs} provides a straightforward and consistent
generalization of the structures found in the four-parton case to $n$
partons. At the three-loop level, if four-parton correlations arise, they
contribute to the anomalous dimension matrix through a sum over
color `quadrupoles' of the form~(\ref{Delta_n_legs}).
At higher loops, of course, structures that directly correlate the colors
and momenta of more than four partons may also arise.


\section{Conclusions\label{sec:conc}}

Building upon the factorization properties of massless scattering
amplitudes in the soft and collinear limits, recent
work~\cite{Gardi:2009qi,Becher:2009qa} determined the principal
structure of soft singularities in multi-leg amplitudes.
It is now established that the cusp anomalous dimension $\gamma_K$ controls 
all pairwise interactions amongst the hard partons, to all loops, and for
general~$N_c$. The corresponding contribution to the soft anomalous
dimension takes the elegant form of a sum over color dipoles,
directly correlating color and kinematic degrees of freedom. This
recent work also led to strong constraints on any additional
singularities that may arise, thus opening a range of interesting
questions.

In the present paper we studied multiple constraints on the form of
potential soft singularities that couple directly four hard partons, which
may arise at three loops and beyond.
We focused on potential corrections to the
sum-over-dipoles formula that do not require the presence of higher
Casimir contributions to the cusp anomalous dimension $\gamma_K$.
The basic property of these functions is that they satisfy the
homogeneous set of differential
equations~(\ref{Delta_oureq_reformulated}), and therefore they can be
written in terms of conformally invariant cross
ratios~\cite{Gardi:2009qi}.

Our main conclusion is that indeed, potential structures of this kind
may arise starting at three loops. Their functional dependence on
both color and kinematic variables is, however, severely constrained by
\begin{itemize}
\item{} Bose symmetry;
\item{} Sudakov factorization and momentum-rescaling symmetry,
dictating that corrections must be 
functions of conformally invariant cross ratios;
\item{} collinear limits, in which the (expected) universal properties 
of the splitting amplitude force corrections to vanish (for $n = 4$ 
partons) or be smooth (for $n > 4$ partons) in these limits;
\item{} transcendentality, a bound on which is expected to be 
saturated at three loops, based on the properties of ${\cal N}= 4$ 
super-Yang-Mills theory.
\end{itemize}
In the three-loop case, assuming purely logarithmic dependence on the
cross ratios, these constraints combine to exclude all but one
specific structure. The three-loop result for $\Delta_n$ can therefore
be written in terms of the expression $\Delta_4^{(122)}$ in 
\eqn{Delta_case122}, up to an overall numerical coefficient.
Because this structure has the maximal possible
transcendentality, $\tau=5$, its coefficient is a
rational number. For all we know now, however, this
coefficient may vanish. It remains for future work to decide
whether this contribution is present or not.
Considering also polylogarithmic functions of conformally
invariant cross ratios in $\Delta_4$, we find that at three loops
at least two additional acceptable functional forms arise, 
\eqns{Delta_case122_mod}{Delta_case311_mod}.

The range of admissible functions at four loops is even larger. 
A particularly interesting feature at this order is the possible
appearance of contributions proportional to quartic Casimir
operators, not only in the cusp anomalous dimension, but in 
four-parton correlations as well. 

Explicit computations at three and four loops will probably be 
necessary to take the next steps toward a complete understanding of
soft singularities in massless gauge theories.


\vskip0.8cm
{\bf Acknowledgments}
\vskip0.2cm

We thank Thomas Binoth, David Kosower and George Sterman for stimulating 
discussions. We thank Thomas Becher and Matthias Neubert for a suggestion
leading to a streamlined analysis in \sect{sec:SA}.
L.~D. and L.~M. thank CERN for hospitality while this work was completed.
This research was supported by the US Department of Energy under contract
DE--AC02--76SF00515, by MIUR (Italy) under contract 2006020509$\_$004,  
and by the European Community's Marie-Curie Research Training Network 
`Tools and Precision Calculations for Physics Discoveries at 
Colliders'  (`HEPTOOLS'), under contract MRTN-CT-2006-035505.


\vskip1.2cm

\appendix 

\section{The four-parton amplitude with no momentum
recoil\label{sec:mom_conserv}}

Here we investigate briefly the simplifications of the potential
forms for $\Delta_4$ at three loops that result from using
momentum-conservation relations special to the four-parton amplitude,
$p_1+p_2+p_3+p_4=0$.
Thus we exclude here the presence in the amplitude of other
colorless particles that might carry recoil momentum.
As before, all momenta are light-like, $p_i^2=0$, so
the momentum invariants are now related by
\[
p_1\cdot p_2 + p_2\cdot p_3 + p_1\cdot p_3\,=\,0\,,
\]
as well as $p_3\cdot p_4 = p_1\cdot p_2$, {\it etc.}\ \  
Using this relation, all three cross ratios entering 
$\Delta_4$ can be expressed in terms of a single dimensionless ratio, 
\begin{equation}
\label{r_def}
r \equiv \frac{p_2\cdot p_3}{p_1\cdot p_2}
\, = \,- \, 1 - \frac{p_1\cdot p_3}{p_1\cdot p_2} \,.
\end{equation}
Substituting into \eqn{rhoijkl_mod}, we have 
\begin{align}
\label{rhoijkl_r}
\begin{split}
L_{1234} \, &= \, \,\ln\left(\left|\frac{1}{1+r}\right|^2 
\,{\rm e}^{-{\rm i}\pi
(\lambda_{12}+\lambda_{34}-\lambda_{13}-\lambda_{24}) }
 \right) \, ,
\\
L_{1423} \, &= \, \,\ln\left(\left|r\right|^2
\,{\rm e}^{-{\rm i}\pi
(\lambda_{14}+\lambda_{23}-\lambda_{12}-\lambda_{34}) }
 \right) \, ,
\\
L_{1342} \, &= \, \,\ln\left(\left|\frac{1+r}{r}\right|^2
\,{\rm e}^{-{\rm i}\pi
(\lambda_{13}+\lambda_{24}-\lambda_{14}-\lambda_{23}) }
\right) \, ,
\end{split}
\end{align}
where we used the variables $\lambda_{ij}$, defined below \eqn{rhoij},
to keep track of the unitarity phases.

\begin{figure}[htb]
\begin{center}
\includegraphics[angle=0,width=13.8cm]{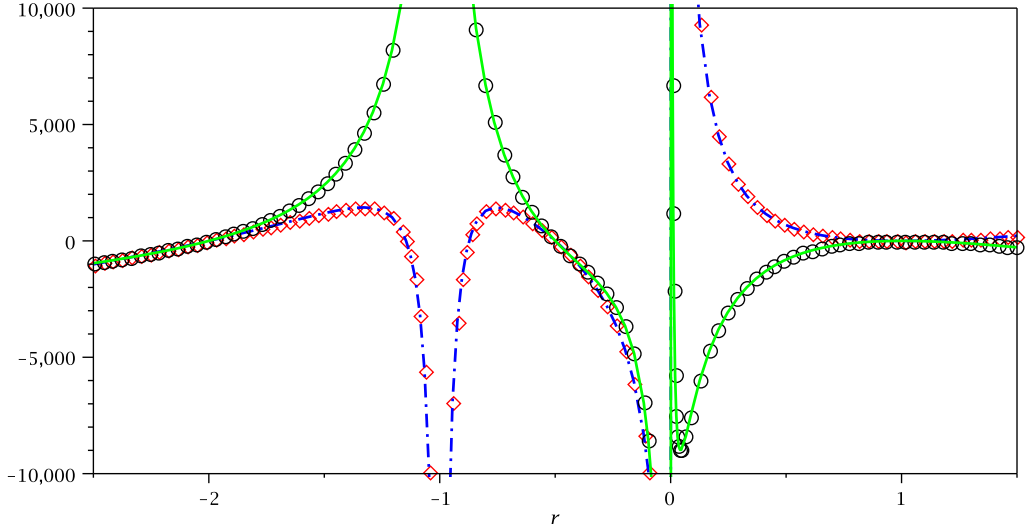}
\caption{The logarithmic functions $\Delta_4$ with transcendentality
$\tau = 5$ introduced in \eqns{Delta_case122}{Delta_case311},
respectively, in the case of a four-parton
amplitude with no recoil. In each case we separately display the real
(dot-dash) and imaginary (solid line) parts as a function of the
ratio $r$, defined in \eqn{r_def}, multiplying the color coefficient 
$f^{ade} f^{cbe}$ after the Jacobi identity has been taken into
account, as in \eqn{Delta_case122_one_variable}. In each case symbols
represent \eqn{Delta_case122} while the lines stand for
\eqn{Delta_case311}.  The plot demonstrates that the two are identical
once the Jacobi identity is taken into account. 
\label{4legs_no_recoil_functions_log}}
\end{center}
\end{figure}


Let us now examine the behavior of the three expressions for $\Delta_4$ 
that we found admissible, as functions of $r$.  As mentioned in 
\sect{CollimitSubsection}, in the four-parton case with no momentum recoil,
$p_i\cdot p_j$ is not a collinear limit, but a forward or backward scattering
limit.  There are three channels to consider:
\begin{itemize}
\item[a)]{} 
$p_1$ and $p_2$ incoming, $p_3$ and $p_4$ outgoing.
The physical region is $- 1 < r < 0$.
\item[b)]{} 
$p_1$ and $p_3$ incoming, $p_2$ and $p_4$ outgoing.
The physical region is $0 < r < \infty$.
\item[c)]{} 
$p_1$ and $p_4$ incoming, $p_2$ and $p_3$ outgoing.
The physical region is $-\infty < r < -1$.
\end{itemize}
The two endpoints of each physical interval are the forward and
backward scattering limits.  Using \eqn{rhoijkl_r}, we can read
off the phases associated with each of the logarithms of the
cross ratios in these three physical regions. The results are
summarized in Table~\ref{table:analytic_continuation_L_ijkl}.
\begin{table}
\begin{center}
\begin{tabular}{|l|l|l|l|}
\hline
&  a) \, $- 1 < r < 0$  & b)\, $0 < r < \infty$  & c)\, $-\infty < r < -1$ \\
\hline
 $L_{1234}/2$    & $-\ln |1+r| -{\rm i}\pi$ 
&   $-\ln |1+r| +{\rm i}\pi$   &      $-\ln |1+r|$ \\
 $L_{1423}/2$    & $\ln|r|+{\rm i}\pi$
&   $\ln|r|$                    &      $\ln|r|-{\rm i}\pi$ \\
 $L_{1342}/2$    & $ \ln|(1+r)/r|$
&   $\ln|(1+r)/r|-{\rm i}\pi$    &    $\ln|(1+r)/r|+{\rm i}\pi$ \\
\hline
\end{tabular}
\end{center}
\caption{Analytic continuation of the three cross ratios into the
three physical regions in a $2\to 2$ scattering amplitude.
\label{table:analytic_continuation_L_ijkl}}
\end{table}

We have seen that, due to the Jacobi identity, the three terms in
$\Delta_4$ corresponding to antisymmetrization of any pair of
color indices are related to each other, leaving just two
independent terms. For example, in \eqn{Delta_case122} we can
substitute $f^{bae}f^{cde} = - f^{ade}f^{cbe} - f^{cae}f^{dbe}$, 
getting
\begin{align}
\label{Delta_case122_one_variable}
\begin{split}
\Delta_4^{(122)}(\rho_{ijkl}) & = \, 
\, {\bf T}_1^{a} {\bf T}_2^{b} {\bf T}_3^{c} {\bf T}_4^{d} \,
\,\,L_{1234}\,L_{1423}\,L_{1342}
\\
&\Bigg\{
f^{ade}f^{cbe}\,L_{1423} \Big[\,L_{1342}\,-\,  L_{1234}\Big]+
f^{cae}f^{dbe}\,L_{1234}\,\Big[ L_{1342}\,-\,L_{1423} \Big]
\Bigg\}\,.
\end{split}
\end{align}
The resulting dependence on $r$ is shown in
Figure~\ref{4legs_no_recoil_functions_log}, which displays the
coefficient of $f^{ade}f^{cbe}$, after using the Jacobi identity,
as in \eqn{Delta_case122_one_variable}. The plot shows that the
result is the same when starting with either  \eqn{Delta_case122}
or \eqn{Delta_case311}, as proven in (\ref{122_and_311_combination}).
In a similar way one can use the Jacobi identity for the other
admissible functions given in  eqs.~(\ref{Delta_case122_mod})
and (\ref{Delta_case311_mod}). Of course, each of them yields a
different function of $r$.


\end{document}